\documentclass[%
 reprint,
superscriptaddress,
longbibliography,
 amsmath,amssymb,
 aps,prx,
]{revtex4-2}

\usepackage{graphicx}
\usepackage{dcolumn}
\usepackage{bm}
\usepackage{hyperref}
\usepackage{bbold}
\usepackage{makecell,hhline}
\usepackage{amsmath,mathtools,amssymb}
\usepackage{xcolor}
\usepackage{soul}

\newcommand{\mK}{\mathsf{K}}
\newcommand{\mW}{\mathsf{W}}
\newcommand{\mM}{\mathsf{M}}
\newcommand{\mJ}{\mathsf{J}}
\renewcommand{\vec}[1]{\boldsymbol{#1}}
\newcommand{\vv}{\vec{v}}

\newcommand{\vp}{\vec{p}}

\newcommand{\vF}{\vec{F}}

\newcommand{\la}{\langle}
\newcommand{\ra}{\rangle}

\newcommand{\ignore}[1]{{}}
\newcommand{\nobibentry}[1]{{\let\nocite\ignore\bibentry{#1}}}

\newcommand{\bibfnamefont}[1]{#1}
\newcommand{\bibnamefont}[1]{#1}

\begin{document}

\title{Probe thermometry with continuous measurements}

\author{Julia Boeyens}
 \email{julia.boeyens@uni-siegen.de}
\affiliation{Naturwissenschaftlich-Technische Fakultät, Universität Siegen, Siegen 57068, Germany}

\author{Björn Annby-Andersson}
\affiliation{Physics Department and NanoLund, Lund University, Box 118, 22100 Lund, Sweden}

\author{Pharnam Bakhshinezhad}
\affiliation{Technische Universität Wien, Stadionallee 2, 1020 Vienna, Austria}
\affiliation{Physics Department and NanoLund, Lund University, Box 118, 22100 Lund, Sweden}

\author{Géraldine Haack}
\affiliation{Département de Physique Appliquée, Université de Genève, 1211 Genève, Switzerland}

\author{Martí Perarnau-Llobet}
\affiliation{Département de Physique Appliquée, Université de Genève, 1211 Genève, Switzerland}

\author{Stefan Nimmrichter}
\affiliation{Naturwissenschaftlich-Technische Fakultät, Universität Siegen, Siegen 57068, Germany}

\author{Patrick P. Potts}%
\affiliation{Department of Physics,
University of Basel and Swiss Nanoscience Institute, Klingelbergstrasse 82, 4056 Basel, Switzerland}%

\author{Mohammad Mehboudi}
\email{mohammad.mehboudi@tuwien.ac.at}
\affiliation{Technische Universität Wien, Stadionallee 2, 1020 Vienna, Austria}
\affiliation{Département de Physique Appliquée, Université de Genève, 1211 Genève, Switzerland}

\date{\today}

\begin{abstract}
Temperature estimation plays a vital role across natural sciences. A standard approach is provided by probe thermometry, where a probe is brought into contact with the sample and examined after a certain amount of time has passed. In many situations however, continuously monitoring the probe may be preferred. Here, we consider a minimal model, where the probe is provided by a two-level system coupled to a thermal reservoir. Monitoring thermally activated transitions enables real-time estimation of temperature with increasing accuracy over time. 
Within this framework we comprehensively investigate thermometry in both bosonic and fermionic environments employing a Bayesian approach. 
Furthermore, we explore adaptive strategies and find a significant improvement on the precision. Additionally, we examine the impact of noise and find that adaptive strategies may suffer more than non-adaptive ones for short observation times. 
While our main focus is on thermometry, our results are easily extended to the estimation of other environmental parameters, such as chemical potentials and transition rates.
\end{abstract}

\maketitle


\section{Introduction}

Temperature plays a prominent role in the study of physical systems, as it is arguably the most relevant state parameter that determines their behaviour. Therefore, thermometry is often a preliminary step in quantum experiments. However, it also introduces measurement disturbance~\cite{RevModPhys.86.1261} in exchange for precision. The theory of quantum metrology~\cite{giovannetti2011advances,paris2009quantum,Toth_2014} identifies optimal precision-disturbance trade-offs in parameter estimation with quantum resources, and quantum thermometry is the branch specialised to temperature estimation~\cite{Mehboudi_2019,DePasquale2018}.

A common framework to study the precision limits in thermometry is to consider a probe coupled to the sample of interest at temperature $T$. Over time, the probe gains information about the sample's temperature, which is later accessed through direct measurements on the probe. 
Relevant experimental realisation of probe thermometry include single-atom  probes for ultracold gases~\cite{Hohmann2016,Bouton2020,Adam2022}, NV centres acting as thermometers of living cells \cite{kucsko2013nanometre,fujiwara2020real}, and nanoscale electron calorimeters  \cite{Gasparinetti2015,halbertal2016nanoscale,karimi2020reaching}. Theoretically, much progress has been achieved on characterising the fundamental precision limits of probe thermometry in frequentist and Bayesian approaches~\cite{PhysRevLett.114.220405,Paris_2016,de2016local,PhysRevLett.127.190402,PhysRevLett.128.130502,PhysRevA.104.052214,PhysRevA.105.012212}, the precision scaling at ultralow temperatures~\cite{Potts2019fundamentallimits,PhysRevB.98.045101,PhysRevResearch.2.033394}, the impact of strong coupling and correlations~\cite{PhysRevA.96.062103,PhysRevLett.122.030403,PhysRevLett.128.040502,PhysRevResearch.4.023191,brenes2023multi}, measurement back action~\cite{albarelli2023invasiveness,PhysRevA.97.032129}, as well as enhanced sensing via non-equilibrium probes~\cite{PhysRevA.82.011611,sabin2014impurities,PhysRevLett.125.080402,razavian2019quantum,RAZAVIAN2019825,PhysRevA.91.012331,Sekatski2022optimal,PhysRevLett.123.180602,mirkhalaf2022operational,brattegard2023thermometry}. While providing remarkable progress on our understanding of thermometry, previous works are based on the assumption that the probe is measured and subsequently reset or discarded. In this work, we depart from this paradigm and consider thermometry through continuous measurements of the probe, where information on the sample's temperature is continuously extracted while the probe is interacting with it. Total measurement time is thus the major resource here, and there is no hidden time cost for probe preparation, as often ignored in previous works. This scenario has to date received little attention, notable exceptions are provided by Refs.~\cite{smiga2023stochastic,Radaelli_2023}. Another reason to pursue this scenario is the remarkable experimental progress on continuous measurements including charge measurements \cite{Lu_2003,Gustavsson_2006,Arnold_2014}, homo- and heterodyne detection \cite{ZHANG20171,wiseman_milburn_2009}, and magnetometry~\cite{sayrin2011real,vijay2012stabilizing,physRevLett.104.013601,PhysRevLett.120.040503,physRevLett.104.013601}.

With this aim, we consider a minimal model where a two-level probe weakly interacts with a thermal bath while being continuously monitored---see Fig.~\ref{fig:sketch}. Depending on the nature of the bath, either bosonic or fermionic, this can correspond to a superconducting qubit coupled to the electromagnetic environment or a quantum dot coupled to an electronic reservoir. We construct temperature estimators and characterize their estimation errors, employing a Bayesian approach \cite{PhysRevLett.127.190402,PhysRevLett.128.130502,PhysRevA.104.052214,PhysRevA.105.012212}. In the long-time limit (or, the large-data limit), the Fisher information, which can be given analytically, determines bounds on these errors. 
We dicuss the saturability of these bounds via non-adaptive and adaptive measurement strategies, where the energy gap of the probe is tuned during the protocol via a suitable feedback. Finally, we characterize the robustness of our results to noisy measurements and a finite bandwidth of the detector,  bringing our considerations closer to experimental platforms.  

The paper is structured as follows: In Section~\ref{sec:system}, we introduce a Markov jump process to describe the system trajectories subject to  a bosonic or fermionic bath.
In Section~\ref{sec:frequentist}, we discuss thermometry in the large-data limit and present an analytical expression for the Fisher information of the trajectories---Eq.~\eqref{eq:fisheralltimes}. This can be used not only for thermometry, but also for estimating other parameters such as the thermalisation rate. Section~\ref{sec:BayesanThermometry} is devoted to the Bayesian approach to thermometry. After defining a relative error quantifier and the optimal estimator that minimises it on average, we present the tight Bayesian Cram\'er-Rao bound in Eq.~\eqref{eq:TBCRB} which we then use to implement adaptive feedback on the probe to improve precision. In Section~\ref{sec:noisy}, we explore the impact of measurement noise on the quality of our continuously monitored thermometer. Finally, in Section~\ref{sec:conclusions}, we conclude and discuss some future directions. 

\section{System and trajectories}\label{sec:system}
\begin{figure}
    \centering
    \includegraphics[width=\linewidth]{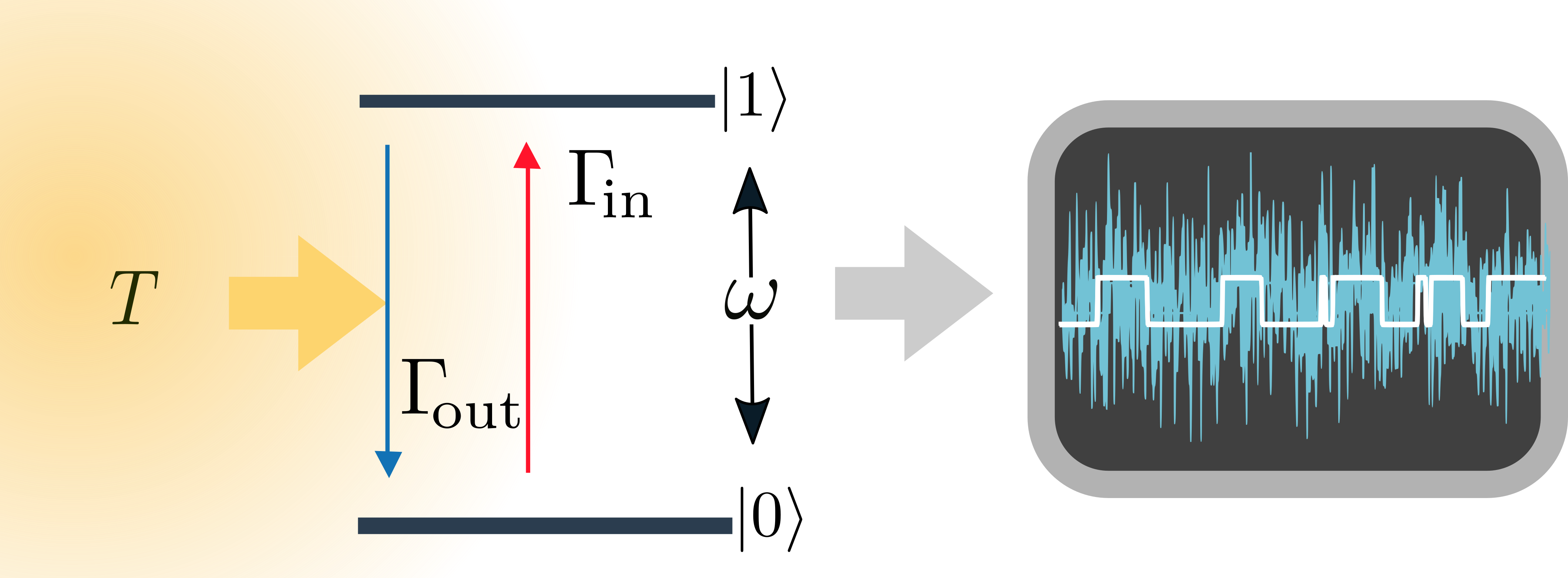}
    \caption{Sketch of continuous thermometry. A two-level system interacts with a bath of unknown temperature $T$ with coupling rates $\Gamma_{\rm in}$ and $\Gamma_{\rm out}$ into and out of the excited state. The energy gap $\omega$ of the system can be changed to improve the performance of the thermometer. The state of the two-level system evolves on a telegraph-like trajectory (white line) and is monitored continuously by a weak measurement, which results in a noisy signal with finite bandwidth (blue). }
    \label{fig:sketch}
\end{figure}
We consider a two-level system, such as a qubit or a quantum dot, coupled to a thermal bath (see Fig.~\ref{fig:sketch}). The two-level system provides the probe which is used to determine the temperature of the sample, provided by the thermal bath. Denoting the two states of the probe by $0$ and $1$, the system dynamics is described by the rate equation
\begin{equation}
    \label{eq:rateeq}
\partial_t \begin{pmatrix}
p_0(t) \\ p_1(t)
\end{pmatrix} = \begin{pmatrix}
-\Gamma_{\rm in} && \Gamma_{\rm out}\\
\Gamma_{\rm in} && -\Gamma_{\rm out}
\end{pmatrix}
\begin{pmatrix}
p_0(t) \\ p_1(t)
\end{pmatrix}.
\end{equation}
Here, $p_j(t)$ denotes the probability that the system is in state $j$ and $\Gamma_{\rm in (out)}$ denotes the rate of the transition $0\rightarrow 1$ ($1\rightarrow 0$). 
It is straightforward to solve Eq.~\eqref{eq:rateeq}, making use of $p_0(t)=1-p_1(t)$, resulting in
\begin{equation}
    \label{eq:solrateeq}
    p_1(t) =  e^{-(\Gamma_{\rm in}+\Gamma_{\rm out})t}\left[p_1-\frac{\Gamma_{\rm in}}{\Gamma_{\rm in}+\Gamma_{\rm out}}\right]+\frac{\Gamma_{\rm in}}{\Gamma_{\rm in}+\Gamma_{\rm out}},
\end{equation}
where $p_1\equiv p_1(0)$. 

The rate equation \eqref{eq:rateeq} describes stochastic jumps between the states $0$ and $1$. Observing these jumps in real time provides information on temperature, because the jump rates are temperature dependent. We note that, while we consider temperature here, any parameter that the rates depend on may be estimated analogously, e.g., properties of the bath spectral density \cite{benedetti:2018}.

Observing the stochastic jumps in the time interval $[0,\tau]$ results in a trajectory $\nu_\tau = \{n(t)|t\in[0,\tau]\}$, where $n(t) \in \{0,1\}$ denotes the occupation of the state at time $t$. The probability density to observe a trajectory is given by
\begin{equation}
    \label{eq:probtraj}
    \rho(\nu_\tau|T) = p_{n_0} \Gamma_{\rm in}^k\Gamma_{\rm out}^l e^{-\Gamma_{\rm in}(\tau-\tau_1)-\Gamma_{\rm out}\tau_1},
\end{equation}
where $p_{n_0}$ denotes the probability of the system being in state $n_0\equiv n(0)$ at $t=0$, $\tau_1$ denotes the total time the system spends in state $1$ along the trajectory, and $k$ ($l$) denotes the number of jumps from state $0\rightarrow 1$ ($1\rightarrow0$) along the trajectory. In Appendix~\ref{app:traj} we derive Eq.~\eqref{eq:probtraj} and we show how $\tau_1$, $k$, and $l$ may be obtained from $n(t)$. We note that because $t$ is a continuous variable, $\rho(\nu_\tau|T)$ is a probability density with a unit that depends on the number of jumps that occur along $\nu_\tau$. The unitless probability to observe a trajectory where the jumps occur within time-windows of width $dt$ is given by $\rho(\nu_\tau|T)(dt)^{l+k}$.

While the above holds true for any pair of rates $\Gamma_{\rm in}$ and $\Gamma_{\rm out}$, we focus here on rates that describe the exchange of energy (and possibly particles) with a thermal bath. Such rates obey a detailed balance relation
\begin{equation}
    \label{eq:detbal}
    \frac{\Gamma_{\rm in}}{\Gamma_{\rm out}} = e^{-\beta \omega},
\end{equation}
where $\beta = 1/(k_BT)$ relates to the inverse temperature, with $k_B$ being the Boltzman constant which we set to $1$ in this work, and $\omega > 0$ is the energy gap between states $0$ and $1$ (with $0$ denoting the ground state). For a bath exchanging particles, we may set the chemical potential to zero without loss of generality and use the same relation.

For concreteness, we consider two widely-used expressions for the rates. The first describes a bosonic thermal bath
\begin{equation}
    \label{eq:bosonicrates}
    \Gamma_{\rm in} = n_B(\omega) \kappa(\omega),\hspace{.5cm}\Gamma_{\rm out} = [n_B(\omega)+1] \kappa(\omega),
\end{equation}
where we introduced the Bose-Einstein distribution 
\begin{equation}
    \label{eq:bed}
    n_B(\omega) = \frac{1}{e^{\beta \omega}-1}.
\end{equation}
With these rates, the considered scenario corresponds, e.g., to a superconducting qubit coupled to an electro-magnetic environment. We will mainly focus on an Ohmic spectral density with $\omega$ well below the cut-off frequency such that $\kappa(\omega) = \kappa' \omega$.

The second expression we use for the rates describes a fermionic bath, describing, e.g., a quantum dot weakly coupled to an electronic reservoir
\begin{equation}
    \label{eq:fermionicrates}
    \Gamma_{\rm in} = n_F(\omega) \Gamma,\hspace{.5cm}\Gamma_{\rm out} = [1-n_F(\omega)] \Gamma,
\end{equation}
where we introduced the Fermi-Dirac distribution
\begin{equation}
    \label{eq:fdd}
    n_F(\omega) = \frac{1}{e^{\beta \omega}+1}.
\end{equation}
For the fermionic rates, we consider a flat spectral density, such that $\Gamma$ is independent of frequency.

\section{Fisher Information}\label{sec:frequentist}
Consider $T^*$ to be the true temperature of the sample which we want to infer by analysing the measured data. 
In thermometry, we want to build an estimator ${\tilde T}(\nu_\tau)$ that maps an observed trajectory $\nu_\tau$ into the best estimate for the temperature. 
To this end, we need to quantify the accuracy of the estimate in terms of a suitable cost (or error) function. An appropriate choice that does not depend on the absolute scale of the true underlying temperature~\cite{Rubio2022} is the relative square distance
\begin{align}\label{eq:relMSE_true}
    D_{\rm R, T^{*}} [{\tilde T},\nu_\tau] \coloneqq \left(\frac{{\tilde T}(\nu_\tau) - T^*}{T^*}\right)^2.
\end{align}
Accordingly, we can quantify the overall performance of the temperature estimation protocol by averaging the relative square distance over all possible trajectories at a particular $T^*$
\begin{align}\label{eq:freq_rel_err}
    D_{\rm R, T^{*}} [{\tilde T}] \coloneqq \int d\nu_{\tau}\rho(\nu_{\tau}|T^*) \left(\frac{{\tilde T}(\nu_\tau) - T^*}{T^*}\right)^2.
\end{align}
Note that this \textit{true} relative mean-square distance is not available in an actual experiment where $T^*$ is unknown. In the frequentist approach to thermometry, one could instead quantify the uncertainty in the estimate of the temperature by, e.g., a confidence region \cite{hoel1984}.

The true relative distance is lower-bounded by the Cram\'er-Rao inequality~\cite{Cramer:107581,Rao1992}. In particular, for any unbiased estimator ${\tilde T}_{\rm u.b.}$ that satisfies
\begin{align}
    \int d\nu_\tau\rho(\nu_{\tau}|T^*){\tilde T}_{\rm u.b.}(\nu_\tau) = T^*,
\end{align}
the true relative distance is lower bounded by
\begin{align}\label{eq:QCRB_Freq}
    D_{\rm R, T^*}^{-1} [{\tilde T}_{\rm u.b.}] \leqslant T^2 F[\rho(\nu_\tau|T)].
\end{align}
Here, $F[\rho(\nu_\tau|T)]$ is the Fisher information of the probability distribution $\rho(\nu_\tau|T)$ with respect to temperature and reads
\begin{equation}
    \label{eq:FT}
    F[\rho(\nu_\tau|T)] = \int d\nu_\tau\rho(\nu_\tau|T) [\partial_T\ln\rho(\nu_\tau|T)]^2.
\end{equation}
For the trajectories described in Eq.~\eqref{eq:probtraj}, we find the Fisher information (see Ref.~\cite{smiga2023stochastic} as well as App.~\ref{app:fisher} for a derivation)
\begin{widetext}
\begin{equation}
    \label{eq:fisheralltimes}
    \begin{aligned}
   & F[\rho(\nu_\tau|T)] = F[p_{n_0}] +\left[p_0 \frac{(\Gamma_{\rm in}')^2}{\Gamma_{\rm in}}+p_1 \frac{(\Gamma_{\rm out}')^2}{\Gamma_{\rm out}}\right]\frac{1-e^{-(\Gamma_{\rm in}+\Gamma_{\rm out})\tau}}{\Gamma_{\rm in}+\Gamma_{\rm out}}\\&+\frac{\Gamma_{\rm in}\Gamma_{\rm out}}{\Gamma_{\rm in}+\Gamma_{\rm out}}\left[\left(\frac{\Gamma_{\rm in}'}{\Gamma_{\rm in}}\right)^2+\left(\frac{\Gamma_{\rm out}'}{\Gamma_{\rm out}}\right)^2\right]\left[\tau-\frac{1-e^{-(\Gamma_{\rm in}+\Gamma_{\rm out})\tau}}{\Gamma_{\rm in}+\Gamma_{\rm out}}\right],
    \end{aligned}
\end{equation}
\end{widetext}
where the prime denotes a derivative with respect to temperature. We stress that while we focus on temperature here, Eq.~\eqref{eq:fisheralltimes} may be applied to any parameter encoded in the rates of a two-level system.
In the long time limit, we may drop all terms that do not grow with the total time of the trajectory $\tau$, which results in
\begin{equation}
    \label{eq:Fisherrates}
    F[\rho(\nu_\tau|T)] = \frac{\Gamma_{\rm in}\Gamma_{\rm out}}{\Gamma_{\rm in}+\Gamma_{\rm out}}\tau\left[\left(\frac{\Gamma_{\rm in}'}{\Gamma_{\rm in}}\right)^2+\left(\frac{\Gamma_{\rm out}'}{\Gamma_{\rm out}}\right)^2\right].
\end{equation}
For a bosonic bath, c.f.~Eq.~\eqref{eq:bosonicrates}, this expression reduces to
\begin{equation}
    \label{eq:fisherbos}
    F[\rho(\nu_\tau|T)] =  \kappa(\omega)\tau \frac{\omega^2}{8k_B^2T^4}\frac{\cosh(\beta\omega)}{\sinh^3(\beta\omega/2)\cosh(\beta\omega/2)}.
\end{equation}
For a fermionic bath, c.f.~Eq.~\eqref{eq:fermionicrates}, we find
\begin{equation}
    \label{eq:fisherferm}
    F[\rho(\nu_\tau|T)] = \Gamma \tau \frac{\omega^2}{8k_B^2T^4}\frac{\cosh(\beta\omega)}{\cosh^4(\beta\omega/2)}.
\end{equation}

If the initial state is the steady state, its contribution to the Fisher information reads
\begin{equation}
    \label{eq:initFish}
    F[p_{n_0}] = \frac{(\Gamma_{\rm out}'\Gamma_{\rm in}-\Gamma_{\rm out}\Gamma_{\rm in}')^2}{\Gamma_{\rm out}\Gamma_{\rm in}(\Gamma_{\rm out}+\Gamma_{\rm in})^2}.
\end{equation}
For rates obeying detailed balance, the steady state corresponds to a thermal state with the Fisher information
\begin{equation}
    \label{eq:fisherinith}
    F[p_{n_0}] =\frac{\omega^2}{2k_B^2T^4}\frac{1}{1+\cosh{(\beta\omega})}.
\end{equation}

A universally applicable and commonly employed frequentist estimator is the \textit{maximum likelihood estimator} (ML)~\cite{kay1993fundamentals}.
It is defined as the temperature that maximises the probability \eqref{eq:probtraj} of the observed trajectory, which can be given exactly as
\begin{align}
    &{\tilde T}_{\rm ML}(\nu_\tau) \coloneqq \arg\underset{T}\max ~(\rho(\nu_\tau|T)) = \frac{\omega}{\log[(1+{\tilde n}_{B})/{\tilde n}_{B}]},\\
&{\tilde n}_{B} \coloneqq \frac{k+l - \kappa(\omega) t + \sqrt{[k+l - \kappa(\omega) t]^2 + 4 k \kappa(\omega) t}}{2 \kappa(\omega) t}.
\end{align}
In the large-data limit, the ML becomes unbiased and saturates the Cram\'er-Rao bound \eqref{eq:QCRB_Freq}.

While the Fisher information sets an ultimate bound on temperature estimation in the asymptotic limit, it can also serve as a means to improve the precision of a thermometer in the course of the measurement. In Sec.~\ref{sec:improved_strategies} below, we will design improved Bayesian estimation strategies based on the Fisher information.
\section{Bayesian thermometry}
\label{sec:BayesanThermometry}
We now focus on the Bayesian approach to thermometry and specify our limited a priori knowledge about the temperature in the form of a prior probability density $\rho(T) \geq 0$, $\int dT\,\rho(T) = 1$. 
Bayes' rule prescribes how to update our knowledge according to the observed trajectory~\cite{gelman1995bayesian}
\begin{align}\label{eq:Bayes_rule}
    \rho(T) \mapsto \rho(T|\nu_\tau) = \frac{\rho(\nu_\tau|T)\rho(T)}{\rho(\nu_\tau)},
\end{align}
given the likelihood $\rho(\nu_\tau|T)$ from Eq.~\eqref{eq:probtraj} and the normalisation factor  $\rho(\nu_\tau) \coloneqq \int dT \rho(\nu_\tau|T)\rho(T)$. 
The posterior distribution $\rho(T|\nu_\tau)$ determines our remaining uncertainty about the actual temperature value after observing $\nu_{\tau}$. Specifically, we can quantify the uncertainty (error) of a temperature estimate $\tilde T (\nu_\tau)$ in a similar manner as before by taking the average of the relative square distance over the posterior
\begin{align}\label{eq:RMSD_given_traj}
    {\rm D}_{\rm R}[{\tilde T},\nu_{\tau}] \coloneqq \int dT \rho(T|\nu_\tau)\left(\frac{{\tilde T}(\nu_\tau)-T}{T}\right)^2.
\end{align}
This represents the \textit{presumed} relative error on temperature an experimentalist would report after recording the trajectory $\nu_{\tau}$. We average the presumed relative error over all possible trajectories to get a trajectory-independent figure of merit for the quality of a given estimator $\tilde T$
\begin{align}\label{eq:RMSD}
    {\rm D}_{\rm R}[{\tilde T}] &\coloneqq \int d\nu_{\tau} \rho(\nu_{\tau}){\rm D}_{\rm R}[{\tilde T},\nu_{\tau}].
\end{align}
Note that, even though the true temperature $T^*$ does not explicitly appear in this expression, we do implicitly assume it is restricted by our specified prior $\rho(T)$. In fact, one could arrive at the same expression \eqref{eq:RMSD} from a different perspective: 
Suppose an experimenter observes trajectories $\nu_\tau$ at the fixed, but unknown true temperature $T^*$. Each trajectory occurs with probability $\rho(\nu_{\tau}|T^*)$ and yields the estimate $\tilde T(\nu_\tau)$, the relative deviation of which from the true value is given by \eqref{eq:relMSE_true}. Averaged over many repetitions, the relative deviation is \eqref{eq:freq_rel_err}---unknown to the experimenter, of course. Then, averaging this also over temperatures drawn from the prior distribution, one finds
\begin{align}
    ~ &\int d T^* \rho(T^*) {\rm D}_{{\rm R},T^{*}}[{\tilde T}]\label{eq:freq_av}\\
    = & \int d T^* \rho(T^*)\int d\nu_{\tau}\rho(\nu_{\tau}|T^*) \left(\frac{{\tilde T}(\nu_\tau) - T^*}{T^*}\right)^2\nonumber\\
    = & \int d \nu_{\tau} \rho(\nu_{\tau})\int d\nu_{\tau}\rho(T^*|\nu_{\tau}) \left(\frac{{\tilde T}(\nu_\tau) - T^*}{T^*}\right)^2 
    \nonumber\\
    =&\int d \nu_{\tau} \rho(\nu_{\tau})\int d\nu_{\tau} {\rm D}_{\rm R}[{\tilde T},\nu_{\tau}]
    \equiv {\rm D}_{\rm R}[{\tilde T}]. \nonumber
\end{align} 
This quantity is a functional of the chosen estimator function $\tilde T$, and by minimizing ${\rm D}_{\rm R}[{\tilde T}]$, we obtain the associated optimal estimator as
\cite{PhysRevA.104.052214}
\begin{align}\label{eq:MRMSDE}
    {\tilde T}_{\rm R}(\nu_\tau) \coloneqq \frac{\int dT \rho(T|\nu_\tau)/T}{\int dT \rho(T|\nu_\tau)/T^2}.
\end{align}
Had we chosen a different cost function than \eqref{eq:RMSD_given_traj} to quantify the uncertainty of the temperature estimate, we would have obtained a different optimal estimator \cite{PhysRevLett.127.190402,Rubio2022}.
For example, we obtain a Bayesian estimator that is tightly related to the ML (and often easy to calculate, and more precise than ML in the small-data limit) if we simply maximize the posterior 
\begin{align}
    {\tilde T}_{\rm MP}(\nu_{\tau})\coloneqq \arg\underset{T}{\max}\rho(T|\nu_{\tau}).
\end{align}
For a prior that is flat over the full range of temperatures, this is equivalent to the ML. In contrast, when the temperature is bounded, ${\tilde T}_{\rm MP}(\nu_{\tau})$ adjusts the ML such that it never estimates a temperature outside the prior domain. In our simulations, we work with a flat prior, 
\begin{align}\label{eq:prior}
    \rho(T)=\begin{cases} 
    [{T_{\rm max} - T_{\rm min}}]^{-1}, &~~~T_{\rm min} \leqslant T \leqslant T_{\rm max},\\
    0, & ~~~{\rm otherwise,}
    \end{cases}
\end{align}
for which the maximum-posterior estimator becomes
\begin{align}
    {\tilde T}_{\rm MP}(\nu_{\tau}) = \max\{\min\{{\tilde T}_{\rm ML}(\nu_{\tau}),T_{\max}\},T_{\min}\}.
\end{align}
Our analysis and techniques will work equally for other choices of prior and cost function; see Appendix~\ref{App:Distances&Estimators}, for a comparison to relevant examples.

The Cram\'er-Rao bound on the true relative square deviation \eqref{eq:QCRB_Freq} also bounds Bayesian figures of merit. 
In particular, if we insert an unbiased estimator into  Eq.~\eqref{eq:freq_av}, we obtain the inequality
\begin{align}\label{eq:TBCRB}
     {\rm D}_{\rm R}[{\tilde T}] 
     & = \int dT\rho(T) D_{\rm R,f}[{\tilde T}]
     & \overset{{\tilde T}_{\rm u.b.}}{\geqslant} \int \frac{dT \rho(T)}{T^2 F[\rho(\nu_{\tau}|T)]}
     ,
\end{align}
see also \cite{bacharach2019some,e20090628}. 
Although this bound \eqref{eq:TBCRB}---sometimes referred to as the tight Bayesian Cram\'er-Rao bound---strictly holds only for unbiased estimators, we expect that Bayesian estimators will respect and are able to saturate it as their biases generally vanish in the limit of large data.
For our system, a vanishing bias is ensured by the Bernstein-Von Mises theorem for Markov processes~\cite{borwanker1971bernstein,Bernstein} (see also Appendix~\ref{app:asymptotic} for a simple derivation), which states that in the long-time limit, the posterior $\rho(T|\nu_\tau)$ takes the shape of a Gaussian around the actual temperature, with a variance that is equal to the inverse of the Fisher information.

\section{Improving precision: adaptive vs non-adaptive strategies}\label{sec:improved_strategies}

\subsection{A non-adaptive strategy}

So far we have characterised the trajectories in Eq.~\eqref{eq:probtraj} given that the thermometer was prepared in a state $p_{n_0}$. For long trajectories, this initial state  plays a vanishing role in the estimation error; whereas the value $\omega$ of the thermometer's energy gap is crucial. In a non-adaptive strategy, we may still be free to tune the gap at the beginning, and it is indeed worth tuning it wisely. While a global optimisation which aims at minimising the error for a given total time $\tau$ is complex and may require heavy numerical simulations, we can still design a strategy to tune the gap such that it is optimal at least for long enough times. Our proposed strategy is therefore to tune the gap to one that minimises the error in the long-time limit. That is, we aim at minimising the right hand side of Eq.~\eqref{eq:TBCRB}
    \begin{align}
        \omega^*_{\rm n. ad.} \coloneqq {\arg\underset{\omega}\min}
        \int \frac{dT \rho(T)}{T^2 F[\rho(\nu_{\tau}|T)]}.
    \end{align}
    This optimal value only depends on the prior and other fixed parameters. In Fig.~\ref{fig:Adaptive_vs_non} we depict a Monte-Carlo simulation of the relative error for the optimal estimator [c.f.~Eq.~\eqref{eq:MRMSDE}] in this non-adaptive scenario. As one can see, asymptotically the error approaches the tight Cram\'er-Rao bound \eqref{eq:TBCRB}, i.e., asymptotically this is an optimal strategy.
\subsection{A greedy adaptive strategy}\label{sec:greedy_adaptive} 
\begin{table*}
	\begin{center}
	\begin{tabular}{ |c|c||c|c|c|  }
		\hline
		 & $\text{Strategy}$  & \makecell{The optimal gap\\Example: $T_{\min}=0.1~\epsilon$, $T_{\max}=10~\epsilon$} & \makecell{Optimal value for \\ $T^2F[\rho(\nu_{\tau}|T)]$}   & \makecell{$\int_{T_{\min}}^{T_{\max}} dT \rho(T)\left[T^2F[\rho(\nu_{\tau}|T)]\right]^{-1}$ \\ Example: $T_{\min}=0.1~\epsilon$, $T_{\max}=10~\epsilon$} \\
		\hhline{|=|=|=|=|=|}
        1 & \makecell{Non adaptive with\\ initially optimised gap}   &   $\approx 0.4595~\epsilon$  & --- & $0.4133~ (  \kappa^{\prime}\epsilon~\tau)^{-1}$\\
         \hline
		2 & \makecell{Adaptive\\ asymptotically thermal}   & $          \approx 2.4750 ~T $ & $ 1.5430 ~ \kappa^{\prime } \tau T$ &      $0.3015~(\kappa^{\prime}\epsilon~ \tau)^{-1}$ \\
		\hline
		
	\end{tabular}
\end{center}
\caption{The different strategies proposed in this work and their asymptotic precision for a Bosonic bath---see TABLE. \ref{table:strategies_fermionic} for a fermionic bath. 
(1) The non-adaptive strategy, in which the gap is chosen once and then left unchanged. That is, we choose the gap that minimises $\int \rho(T) dT \left[T^2F[\rho(\nu_{\tau}|T)]\right]^{-1}$ with $F[\rho(\nu_{\tau}|T)]$ from Eq.~\eqref{eq:Fisherrates}. Once the gap is chosen at the begining of the process, it will not be adaptively changed. One can see that for the parameter range considered here, this strategy performs reasonably good compared to the adaptive one.
(2) A practical adaptive strategy that only controls the gap and leaves the state untouched. The gap is chosen such that in the asymptotic limit it is optimal for a thermal state (i.e., the gap that maximises $F[\rho(\nu_{\tau}|T)]$ in Eq.~\eqref{eq:Fisherrates}). 
For the parameters that we consider, the adaptive strategy (2) can outperform the non-adaptive strategy (1) by $\left[{\rm D}_{\rm R}[{\tilde T}_{\rm R}](\rm ad.) - {\rm D}_{\rm R}[{\tilde T}_{\rm R}](\rm n.~ad.)\right]/{{\rm D}_{\rm R}[{\tilde T}_{\rm R}](\rm n.~ad.)}\approx 40\%$---in the asymptotic limit. Here, temperature and energies are expressed in units of $\epsilon$ while $\tau$ has units of $\epsilon^{-1}$. The coupling $\kappa^{\prime}$ is dimensionless. In the examples we have set $T_{\min}=0.1\epsilon$, $T_{\max}=10\epsilon$, $\epsilon=1$, and $\kappa^{\prime}=1$.}
\label{table:strategies}
\end{table*}
\begin{table*}
	\begin{center}
	\begin{tabular}{ |c|c||c|c|c|  }
		\hline
		 & $\text{Strategy}$  & \makecell{The optimal gap\\Example: $T_{\min}=0.1~\epsilon$, $T_{\max}=10~\epsilon$} & \makecell{Optimal value for \\ $T^2F[\rho(\nu_{\tau}|T)]$}   & \makecell{$\int_{T_{\min}}^{T_{\max}} dT \rho(T)\left[T^2F[\rho(\nu_{\tau}|T)]\right]^{-1}$ \\ Example: $T_{\min}=0.1~\epsilon$, $T_{\max}=10~\epsilon$} \\
		\hhline{|=|=|=|=|=|}
        1 & \makecell{Non adaptive with\\ initially optimised gap}   &   $\approx 1.5401~\epsilon$  & --- & $132.79 ~(\Gamma ~ \tau)^{-1}$\\
         \hline
		2 & \makecell{Adaptive\\ asymptotically thermal}   & $          \approx 2.6672 ~ T $ & $ 0.3795 ~ \Gamma \tau$ &      $   2.6350~( \Gamma ~ \tau)^{-1}$ \\
		\hline
		
	\end{tabular}
\end{center}
\caption{Same as TABLE. \ref{table:strategies}, but for a Fermionic bath with flat spectral density (i.e., $s=0$). Evidently, the choice of the strategy is more impactful than the Bosonic bath. Specifically, one can see adaptive strategies can improve the asymptotic precision more than an order of magnitude compared to the non-adaptive one. One should note that, this is mainly stemming from the choice of a flat spectral density, that is more appropriate  for fermionic baths. Again, we express all the parameters in terms of $\epsilon$, such that temperature $T$, frequency $\omega$, and the coupling $\Gamma$ have the $\epsilon$ dimension, while time $\tau$ has the dimension $\epsilon^{-1}$. For this table we have set $T_{\min}=0.1 \epsilon$, $T_{\rm max} = 10 \epsilon$, and $\Gamma = \epsilon=1$. }
\label{table:strategies_fermionic}
\end{table*}
Here, we take a step further and assume that we have the freedom to change the probe's energy gap $\omega$ during the measurement; that is, we can adaptively tune it. Once again, a global optimisation strategy may be costly. Thus, we will seek a simple alternative. 
Again, we design a strategy that is optimal in the long-time limit which works as follows: 
Suppose that at time $t$ we have observed a trajectory $\nu_t$ and estimate the temperature to be ${\tilde T}(\nu_t)$. We
tune the gap to the one that maximises ${\tilde T}^2(\nu_t) F[\rho(\nu_{\tau}|{\tilde T}(\nu_t))]$. Note that $\nu_\tau$ is a variable that is integrated over when determining the Fisher information, see Eq.~\eqref{eq:FT}, not to be mistaken with the observed trajectory $\nu_t$ which is used to build the estimator, see Eq.~\eqref{eq:Fisherrates}. That is, at any time $t$ and after observing the trajectory $\nu_t$ we tune the gap to
\begin{align}
    \omega^{*}_{\rm ad.}(\nu_t) \coloneqq  {\arg\underset{\omega}\max}~{\tilde T}^2(\nu_t) F[\rho(\nu_{\tau}|{\tilde T}(\nu_t))].
\end{align}
Using the expressions for the Bosonic and fermionic Fisher information we have
\begin{align}
    \omega^{*}_{\rm ad.}(\nu_t)   & \approx 2.4750~ {\tilde T}(\nu_t),\hspace{.2cm} \text{Bosonic},\\
    \omega^{*}_{\rm ad.}(\nu_t)   & \approx 2.6672~ {\tilde T}(\nu_t),\hspace{.2cm} \text{Fermionic}.
\end{align}
This strategy guarantees that the gap asymptotically converges to the fixed optimal value, since we expect the estimator to approach the true temperature at long times. The convergence to a fixed gap is important; even if our feedback is delayed, it eventually tunes the gap to the optimal one for the true temperature. 

\begin{figure}
    \centering
    \includegraphics[width=\linewidth]{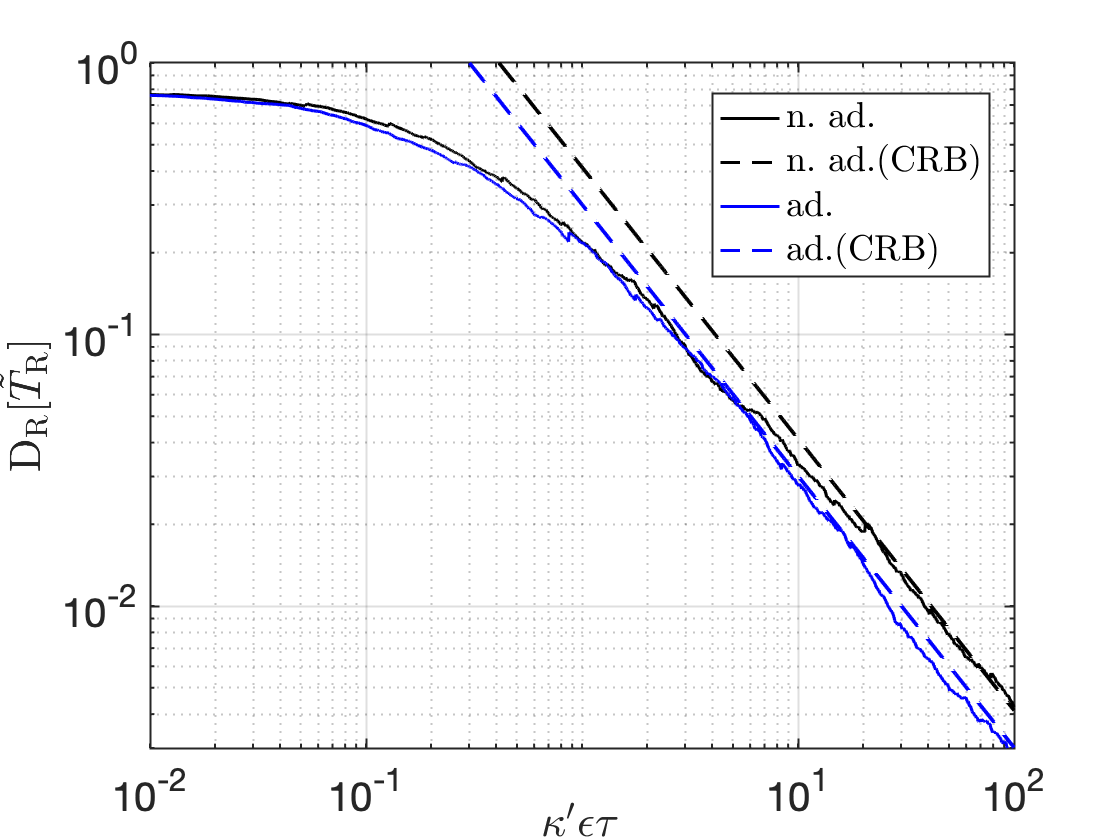}
    \caption{The relative error $D_{\rm R}[{\tilde T}_{\rm R}]$ in thermometry of a bosonic bath as a function of time. Here we evaluate the relative error by a Monte Carlo simulation that approximates Eq.~\eqref{eq:RMSD} by randomly sampling a temperature from the prior according to which we randomly sample a trajectory. We repeat this process $1000$ times and take the average. The error is plotted for both the adaptive (solid blue), and the non adaptive (solid black) scenarios. The CRB lines correspond to the r.h.s. of Eq.~\eqref{eq:TBCRB}. At long times, both strategies approach this asymptotic bound both for adaptive (dashed blue) and non-adaptive (dashed black) scenarios. Our simulations show that the adaptive strategies outperform the non-adaptive ones even in the non-asymptotic times. In the simulations we choose the parameters according to TABLE. \ref{table:strategies}, that is $T_{\min}=0.1\epsilon$, $T_{\max}=10\epsilon$, $\epsilon=1$, and $\kappa^{\prime}=1$. In the adaptive strategy, the frequency changes in each time step. These graphs are obtained by simulating $ 1000 $ trajectories with temperatures that are randomly sampled from the prior distribution. }
    \label{fig:Adaptive_vs_non}
\end{figure}

In Fig.~\eqref{fig:Adaptive_vs_non} we demonstrate how the use of the adaptive strategy can outperform a non-adaptive scenario and saturate the asymptotic CRB at the optimal gap. 

Let us remark that the aforementioned simple strategies are motivated by asymptotic figures of merit. In the transient regime, however, alternative strategies might be more appropriate. 

\section{Noisy measurements}\label{sec:noisy}

Many experimental platforms, e.g., semiconductor quantum dots \cite{Barker-PRL-2022}, include noise in the measured trajectory, rather than the telegraph-like trajectory defined above Eq.~(\ref{eq:probtraj}). Additionally, it is common that the detector is limited by a finite bandwidth, introducing a delay in the readout. In this section, we develop a model for Bayesian parameter estimation including these effects, focusing on Gaussian measurement noise. The model is developed for temperature estimation using the two-level probe system defined in Eq.~(\ref{eq:rateeq}), but can be adapted to any other parameter and to more complicated architectures~\cite{Annby-Andersson-PRL-2022}. We use the model to simulate non-adaptive as well as adaptive temperature estimation.

The noisy trajectory is defined as $\nu_\tau=\{D_t|t\in[0,\tau]\}$, where $D_t$ is the outcome of the detector at time $t$. When the system resides in the ground (excited) state, the detector signal randomly fluctuates around 0 (1). The evolution of the system under such a measurement is 
\begin{align}\label{eq:meas_evol}
\boldsymbol{P}_t(\nu_t) =
\mM (D_t|D_{t-dt})e^{\mW dt} \boldsymbol{P}_{t-dt}(\nu_{t-dt}),
\end{align}
where $\boldsymbol{P}_\tau(\nu_\tau)=\left(p_0(\nu_\tau),p_1(\nu_\tau)\right)^{\rm T}$ is a column vector with $p_j(\nu_\tau)$ the joint probability of occupying state $j\in\{0,1\}$ at time $\tau$ and observing $\nu_\tau$. Here $\mW$ is the rate matrix in Eq.~(\ref{eq:rateeq}), determining the time-evolution of the probe system. The matrix $\mM (D|D')$ describes the measurement of outcome $D$, given that $D'$ was observed in the previous timestep. Following Ref.~\cite{Annby-Andersson-PRL-2022}, we find
\begin{align}
\mM&(D|D') =\sqrt{\frac{2\lambda }{\pi\gamma^2dt}} \\
&\begin{pmatrix}
e^{-\frac{2\lambda}{\gamma^2dt} (D-D'e^{-\gamma dt})^2} & 0 \\
0 & e^{-\frac{2\lambda}{\gamma^2dt} [D-(D'e^{-\gamma dt}+\gamma dt)]^2}
\end{pmatrix}\nonumber,
\end{align}
where $\lambda$ is the strength of measurement and $\gamma$ is the bandwidth of the detector. A strong measurement ($\lambda\gg\gamma$) reduces the noise of the detector, while a weak measurement ($\lambda\ll\gamma$) increases the noise. The bandwidth introduces a delay $1/\gamma$ in the detector and substantially dampens all frequency components larger than $\gamma$. For $\lambda\gg\gamma\to\infty$, the noise and lag vanish, and we recover the telegraph signal with the trajectory given by Eq.~(\ref{eq:probtraj}).

The likelihood of observing the trajectory $\nu_\tau$, given the temperature $T$, is calculated via the inner product
\begin{equation}\label{eq:likelihood}
    \rho(\nu_\tau|T) = (1,1)\cdot\hspace{0.5mm}\boldsymbol{P}_\tau(\nu_\tau).
\end{equation}
The vector on the right-hand side can be calculated iteratively according to Eq.~\ref{eq:meas_evol} using an initial distribution $\boldsymbol{P}_0=(p_0,p_1)^{\rm T}$. It is then used to update the current state of knowledge using Bayes rule [see Eq.~(\ref{eq:Bayes_rule})]. The evolution of the state of knowledge with each increment in the measurement register can be equivalently formulated in the language of continuous signal filtering resulting in the Kushner-Stratonovich equation. This is derived in Appendix~\ref{app:KS_eqn}

\begin{figure}
    \centering
    \includegraphics[width=\linewidth]{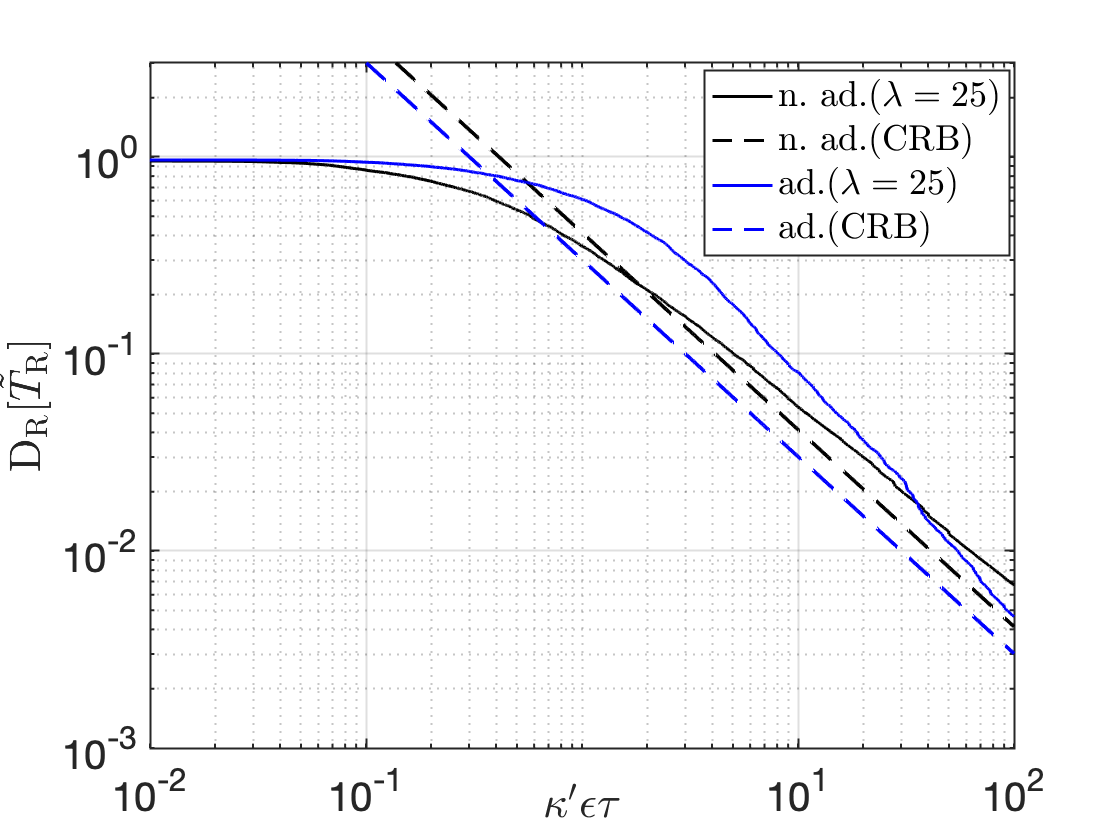}
    \caption{Performance of the temperature estimation protocol when it is limited by a noisy trajectory with a finite bandwidth $\gamma=10.0$. Here, the relative error  is averaged over 1000 trajectories generated at randomly selected temperatures in the range $[0.1,10.0]$ with an initial gap $\omega^*_{\text{n. ad}}$. It is plotted as a function of time and is compared to the adaptive and non-adaptive Cramer-Rao bounds. Increasing the measurement strength $\lambda$ leads to a better accuracy however, the noise prevents the CRB from being reached. For a finite measurement strength, the adaptive strategy performs worse than the non-adaptive strategy at short times. The rest of the parameters are chosen similar to Fig.~\ref{fig:Adaptive_vs_non}, that is $T_{\min}=0.1\epsilon$, $T_{\max}=10\epsilon$, $\epsilon=1$, and $\kappa^{\prime}=1$.}
    \label{fig:noisy_scaling}
\end{figure}
In Fig.~\ref{fig:noisy_scaling}, we show that noisy measurements do not saturate the Cramer-Rao bound Eq.~\eqref{eq:TBCRB}; contrary to what we saw for ideal measurements.
Nonetheless, our adaptive strategy---which is similar to the one we use for the ideal measurement---can reach the Cramer-Rao bound for the non-adaptive ideal measurement at long enough times. Unfortunately, we cannot simulate for arbitrary long times, but the possibility for the adaptive noisy measurement to outperform the non-adaptive ideal measurement is not ruled out. 
Lastly, at 
short times the performance of the adaptive strategy is actually worse than it is for non-adaptive strategies. 
Again, note that our adaptive strategy is not necessarily optimal for the noisy scenario; it is rather designed to be optimal for the ideal measurement and at long times.
In Appendix~\ref{app:bias} we further discuss that a significant estimation bias that persists at low and high temperatures is the reason behind the noisy measurements failing to saturate the Cram\'er-Rao bound. At low temperatures this bias arises from there not being sufficient time for transitions to be observed which results in the estimated temperature being below the actual temperature. This low temperature bias is not affected by the measurement strength. In contrast, at higher temperatures the temperature is overestimated. This is reduced when the measurement strength is increased. Additionally, it shows that one of the ways that the adaptive strategy improves the estimation is by reducing the bias.  In particular, at low temperatures, being able to adjust the gap of the system allows more transitions to occur in the trajectory leading to a better estimate.
\section{Conclusions \& outlook}\label{sec:conclusions}
Our thermometry protocol based on a continuously monitored two-level system offers theoretical tools for temperature estimation (and estimation of other environmental parameters) in quantum systems to various experimental settings dealing with both bosonic and fermionic environments. In particular, our results can be readily exploited in experimental scenarios, e.g., in temperature and chemical potential estimation using continuously monitored quantum dots.

Our results can be related to and contrasted against several other studies in recent years. Indeed, continuous monitoring as a non-destructive method for interrogating quantum systems has found use in parameter estimation tasks particularly in magnetometry~\cite{sayrin2011real,vijay2012stabilizing,physRevLett.104.013601,PhysRevLett.120.040503,physRevLett.104.013601}. Theoretical works have also put forward proposals to surpass the standard quantum limit, and furthermore address the shortcomings in presence of noise~\cite{PhysRevA.69.032109,PhysRevLett.93.173002,PhysRevLett.91.250801,Amoros-Binefa_2021}. On a more fundamental level, the ultimate Bayesian and frequentist bounds have been addressed~\cite{PhysRevA.87.032115,Albarelli_2017,PhysRevA.94.032103,PhysRevLett.112.170401,PhysRevA.90.012330}. These bounds cannot be straightforwardly adopted to our problem since they are either problem-specific or fundamental and thus too generic.
Furthermore, in our case, the parameter to be estimated is a property of the environment, 
which distinguishes our setting from magnetometry and other Hamiltonian estimation tasks. We were able to analytically characterise the trajectories---see Eq.~\eqref{eq:probtraj}---which is crucial in finding the frequentist and Bayesian limits in estimation and designing improved non-adaptive and adaptive protocols.

Our investigation leaves several future directions to proceed. We considered purely classical dynamics, without any quantum coherence. In the presence of coherence, our findings should be revisited to find the optimal measurement and whether quantum correlations are beneficial. Furthermore, we consider a single probe scenario.  
In the presence of multiple interacting (or many-body) probes, quantum correlations may be harnessed for thermometry---similar to the unitarily encoded magnetometry considered in Ref.~\cite{jones2020remote}.     

Finally, our optimal protocols are based on optimising the Cram\'er-Rao bound. They are optimal in the limit of large data (long monitoring time). However, in the finite-data regime, one may be able to design better strategies. This could be done, at a massive computational cost, by minimising the relative distance error numerically. A smarter algorithm that requires less computational resources---at the cost of being sub-optimal---is desirable.

\acknowledgements

P.P.P, M.P.L, and G.H.  acknowledge funding from the Swiss National Science Foundation  (Eccellenza Professorial Fellowship PCEFP2\_194268, Ambizione Grant
No. PZ00P2-186067, PRIMA PR00P2 179748). G.H. also acknowledges NCCR Swissmap. M.M. acknowledges funding from the DFG/FWF Research Unit FOR 2724 ‘Thermal machines in the quantum world’. B.A.A. was supported by the
Swedish Research Council, Grant No. 2018-03921. P.B. is supported by the European Research Council (Consolidator grant ’Cocoquest’ 101043705) and grant number FQXi Grant Number: FQXi-IAF19-07 from the Foundational Questions Institute Fund, a donor advised fund of Silicon Valley Community Foundation.

%

\appendix
\onecolumngrid
\section{Trajectory probabilities}
\label{app:traj}
In this appendix, we provide a derivation of the probability density for trajectories and we show how all quantities appearing in Eq.~\eqref{eq:probtraj} may be expressed through the time-dependent occupation $n(t)$ along a trajectory. 
As a first step, we discretize time in steps of $\delta t = \tau/N$ and define the trajectory as the sequence $\nu_N = [n_j]_{j=0}^N$ with $n_j = n(j\delta t )$.
Since the rate equation in Eq.~\eqref{eq:rateeq} is Markovian, we may write the likelihood of a given trajectory as
\begin{equation}
    \label{eq:probtrajapp1}
     p(\nu_N|T) = p_{n_0}\prod_{j=0}^{N-1} p(n_{j+1}|n_j),
\end{equation}
where $p_{n_0} = p_{n(0)}(t=0)$ denotes the probability for the initial state. The conditional probabilities propagating the state are directly obtained by integrating Eq.~\eqref{eq:rateeq},
\begin{align}
    p(1|0) &= \frac{\Gamma_{\rm in}}{\Gamma_{\rm in} + \Gamma_{\rm out}} \left[ 1 - e^{-(\Gamma_{\rm in}+\Gamma_{\rm out})\delta t} \right] \nonumber \\
    &= n_F (\omega) \left[ 1 - e^{-(\Gamma_{\rm in}+\Gamma_{\rm out})\delta t} \right], \\
    p(0|1) &= \frac{\Gamma_{\rm out}}{\Gamma_{\rm in} + \Gamma_{\rm out}} \left[ 1 - e^{-(\Gamma_{\rm in}+\Gamma_{\rm out})\delta t} \right] \nonumber \\
    &= \left[ 1- n_F (\omega) \right] \left[ 1 - e^{-(\Gamma_{\rm in}+\Gamma_{\rm out})\delta t} \right], \\
    p(0|0) &= 1 - p(1|0), \qquad p(1|1) = 1-p(0|1) .
\end{align}
For our purposes, we will be mainly concerned with continuous trajectories in the limit of $\delta t \to 0$, $\nu_N \to \nu_\tau = n([0,\tau])$. Restricting to finite temperatures at which $(\Gamma_{\rm in} + \Gamma_{\rm out})\delta t \ll 1$, we may expand the propagator probabilities to linear order as
\begin{equation}
    \label{eq:condrates}
    p(1|0) = \delta t \Gamma_{\rm in},\qquad p(0|1) = \delta t \Gamma_{\rm out},
\end{equation}
so that the likelihood \eqref{eq:probtrajapp1} becomes
\begin{equation}
    \label{eq:probtrajapp2}
      p(\nu_\tau|T) = \left(\delta t \Gamma_{\rm in}\right)^k\left(\delta t \Gamma_{\rm out}\right)^l\left(1-\delta t \Gamma_{\rm in}\right)^q\left(1-\delta t \Gamma_{\rm out}\right)^r.
\end{equation}
Here the $k$, $l$, $q$, $r$ denote the numbers of times the corresponding conditional probabilities occur in Eq.~\eqref{eq:probtrajapp1}. We note that in the limit $\delta t \rightarrow 0$, $k,l\ll q,r$ for trajectories with finite probability because the probability for jumps is infinitesimally smaller than the probability for staying in the same state. We may thus interpret $r\delta t=\tau_1$ as the time spent in the state $1$ and similarly for $q\delta t = \tau-\tau_1$. This implies
\begin{equation}
    \label{eq:expid}
    \left(1-\delta t \Gamma_{\rm out}\right)^r = \left(1-\tau_1 \Gamma_{\rm out}/r\right)^r\rightarrow e^{-\Gamma_{\rm out}\tau_1},
\end{equation}
and similarly for the term involving $q$. With these relations, we may recover Eq.~\eqref{eq:probtraj} for the probability density $\rho(\nu_\tau|T) = p(\nu_\tau|T)/(\delta t)^{l+k}$.

While from the above analysis, it should become clear that the quantities $k$, $l$, and $\tau_1$ can be obtained from $n(t)$, we now provide explicit relations in the continuum limit. First, we note that the total time spent in the state $1$ is simply given by the integral
\begin{equation}
    \label{eq:tau1}
    \tau_1 = \int_0^\tau dt \, n(t).
\end{equation}
The variable $k$ denotes the jumps where $n$ changes from $0$ to $1$. It can be expressed as 
\begin{equation}
    \label{eq:k}
    k = \int_0^\tau dn(t) \delta_{dn(t),1}, 
\end{equation}
where we introduced the stochastic increment
\begin{equation}
    \label{eq:dn}
    dn(t) = \lim_{\delta t\rightarrow 0} \left[n(t+\delta t)-n(t)\right].
\end{equation}
Similarly, we find
\begin{equation}
    \label{eq:l}
    l = -\int_0^\tau dn(t) \delta_{dn(t),-1}.
\end{equation}
These expressions imply
\begin{equation}
    \label{eq:k-l}
    k-l = \int_0^\tau dn(t) = n(\tau)-n(0),
\end{equation}
implying that either $k=l$ or $k=l\pm1$.

\section{Fisher information for trajectories}
\label{app:fisher}

Given the likelihood \eqref{eq:probtrajapp1} for Markovian discrete-time trajectories of the probe, the associated Fisher information with respect to temperature reduces to an analytic expression. To see this, we introduce a convenient matrix-vector notation: let the column vector $\vp = (p_0 , p_1)^{\rm T}$ denote the initial probe state, and let $\mK$ be the $2\times2$ propagator with matrix elements $K_{nm} = p(n|m)$. 
Noting that $p(\nu_N | T) = p(n_N|n_{N-1}) p(\nu_{N-1}|T)$ with $p(\nu_0|T) = p_{n_0}$, and introducing the column vector $\vF^{(N)}$ with elements $F^{(N)}_n = \la \delta_{n_N, n} [\partial_T p(\nu_N|T)]^2 \ra$, we find the iterative solution to the Fisher information, 
\begin{align}
    F [p(\nu_N|T)] = (1,1) \vF^{(N)} &= \sum_{n_N} \sum_{\nu_{N-1}} p(n_N | n_{N-1}) p(\nu_{N-1}|T) \left\{ \partial_T \ln \left[ p(n_N | n_{N-1}) p(\nu_{N-1}|T) \right] \right\}^2 \nonumber \\
    &= \sum_{n_N} \sum_{\nu_{N-1}} p(n_N | n_{N-1}) p(\nu_{N-1}|T) \left[ \partial_T \ln p(n_N|n_{N-1}) + \partial_T \ln p(\nu_{N-1}|T) \right]^2 \nonumber \\
    &= (1,1)\mJ \mK^{N-1} \vp + 2 (1,1) (\partial_T \mK) (\partial_T \mK^{N-1}\vp) + (1,1) \mK \vF^{(N-1)} \nonumber \\
    &= (1,1) \mJ \mK^{N-1} \vp + F [p(\nu_{N-1}|T)] = \ldots = (1,1)\mJ \sum_{j=0}^{N-1} \mK^j \vp + F[p_{n_0}]. \label{eq:FI_iterative}
\end{align}
In the third line, we introduce the matrix $\mJ$ with elements $J_{nm} = K_{nm} (\partial_T \ln K_{nm})^2$. In the fourth line, we make use of the fact that $\mK$ is a stochastic matrix with $(1,1)\mK = (1,1)$ and $(1,1)\partial_T \mK = 0$. Finally, we iterate the recursion down to the initial probe state (which may also depend on temperature).

Next, we use that $\mK$ has the two right-eigenvalues $\lambda_1 = 1$ and $\lambda_2 = e^{-(\Gamma_{\rm in} + \Gamma_{\rm out})\delta t}$ with corresponding (unnormalized) right-eigenvectors $\vv_1 = (\Gamma_{\rm out},\Gamma_{\rm in})^{\rm T}$ and $\vv_2 = (1,-1)^{\rm T}$, respectively. This implies 
\begin{equation}
    \mK^j \vp = \frac{\vv_1}{\Gamma_{\rm in} + \Gamma_{\rm out}} 
    + \left( \frac{\Gamma_{\rm in}}{\Gamma_{\rm in}+\Gamma_{\rm out}} - p_1 \right) e^{-(\Gamma_{\rm in} + \Gamma_{\rm out})j \delta t} \vv_2,
\end{equation}
which holds also for $j=0$, as one easily verifies. Inserting this into \eqref{eq:FI_iterative} and carrying out the geometric sum, we get
\begin{equation}
    F [p(\nu_N|T)] = \frac{(1,1)\mJ}{\Gamma_{\rm in}+\Gamma_{\rm out}} \left[ N\vv_1 + \frac{1 - e^{-(\Gamma_{\rm in} + \Gamma_{\rm out})N \delta t}}{1 - e^{-(\Gamma_{\rm in} + \Gamma_{\rm out}) \delta t}}  (\Gamma_{\rm in} p_0-\Gamma_{\rm out}p_1)\vv_2 \right] + F[p_{n_0}], \label{eq:FI_fullSol}
\end{equation}
where the row vector $(1,1)\mJ$ contains the Fisher information of the transition probabilities,
\begin{equation}
    (1,1)\mJ = \left( F[p(n|0)], F[p(n|1)] \right) =  \left( \frac{[\partial_T p(1|0)]^2}{p(1|0)[1-p(1|0)]}, \frac{[\partial_T p(0|1)]^2}{p(0|1)[1-p(0|1)]} \right).
\end{equation}
In the continuum limit $\delta t \to 0$, the transition probabilities are given by \eqref{eq:condrates}, which yields  
\begin{equation}
    \label{eq:fishercond}
    \left( F[p(n|0)], F[p(n|1)] \right) = \left( \left[ \frac{\Gamma_{\rm in}'}{\Gamma_{\rm in}}\right]^2 \Gamma_{\rm in} \delta t,  \left[ \frac{\Gamma_{\rm out}'}{\Gamma_{\rm out}}\right]^2 \Gamma_{\rm out} \delta t \right),
\end{equation}
with the short-hand notation $\Gamma_{\rm in,out}' = \partial_T \Gamma_{\rm in,out}$. Inserting \eqref{eq:fishercond} into the Fisher information expression \eqref{eq:FI_fullSol} and expanding to linear order in $\delta t$ at fixed $\tau = N\delta t$, we arrive at
\begin{align}
    F [p(\nu_\tau|T)] = \frac{\Gamma_{\rm in}\Gamma_{\rm out}\tau}{\Gamma_{\rm in}+\Gamma_{\rm out}} \left[ \left( \frac{\Gamma_{\rm in}'}{\Gamma_{\rm in}}\right)^2 + \left( \frac{\Gamma_{\rm out}'}{\Gamma_{\rm out}}\right)^2 \right]
    + \left[ 1 - e^{-(\Gamma_{\rm in} + \Gamma_{\rm out})\tau} \right] \frac{\Gamma_{\rm in} p_0-\Gamma_{\rm out}p_1}{(\Gamma_{\rm in} +\Gamma_{\rm out})^2} \left[ \frac{(\Gamma_{\rm in}')^2}{\Gamma_{\rm in}} - \frac{(\Gamma_{\rm out}')^2}{\Gamma_{\rm out}} \right]
    + F[p_{n_0}] .
\end{align}
Since the definition of the Fisher information in Eq.~\eqref{eq:FT} implies $F[\rho(\nu_\tau|T)]=F[p(\nu_\tau|T)]$, we recover Eq.~\eqref{eq:fisheralltimes} in the main text.

We remark that, in the opposite limit of $(\Gamma_{\rm in}+\Gamma_{\rm out})\delta t \gg 1$, the probe fully thermalises from one interrogation step to the next, so that the protocol amounts to measuring the initial probe state and $N$ independent probes in thermal equilibrium. Indeed, we find in this regime that $(1,1)\mJ = (1,1) F^{(\rm eq)}$, with $F^{(\rm eq)} = [n_F' (\omega)]^2/n_F(\omega) [1-n_F(\omega)]$ the thermal Fisher information at equilibrium. Hence, \eqref{eq:FI_fullSol} reduces to the expected $ F [p(\nu_N|T)] = N F^{(\rm eq)} + F[p_{n_0}]$.

\section{The asymptotic behaviour of the posterior distribution}\label{app:asymptotic}
Here we provide a simple proof of the asymptotic behaviour, for a more rigerous derivation see~\cite{borwanker1971bernstein}. Let us start by the discrete case. By noting that asymptotically we can see the trajectories as the $N$ outcomes of the population measurement we can write
\begin{align}
    \rho(T|\nu_N) & \propto \rho(T)\prod_{j=0}^{N-1} p(n_{j+1}|n_j,T) = \rho(T)\prod_{j=0}^{1}\prod_{k=0}^{1}  p(j|0,T)^{N_{0\to j}} p(k|1,T) ^{N_{1\to j}}\nonumber\\
    & = \rho(T)\prod_{j=0}^{1}\prod_{k=0}^{1} p(j|0,T)^{N_0 \frac{N_{0\to j}}{N_0}}p(k|1,T)^{N_1 \frac{N_{1\to k}}{N_1}}\nonumber\\
    & \approx \rho(T)\prod_{j=0}^{1}\prod_{k=0}^{1} p(j|0,T)^{N_0 p(j|0,T^*)}p(k|1,T)^{N_1 p(k|1,T^*)},
\end{align} 
where we define $N_0$ the number of times that the state is observed at the ground state, while $N_1=N-N_0$ is the number of times it is observed in the exited state. Furhtermore, we defined $N_{j\to k}$ the number of times that the system starts in the initial state $j$ and ends in the state $k$ and it holds that $N_{j\to 0} + N_{j\to 1} = N_j$. Finally, in the last line we use the fact that, for large enough $N$ one should have $N_{j\to k}/N_j = p(k|j,T^*)$, with $T^*$ being the true temperature that has generated the observed trajectory
\footnote{More rigorously, $N_j [N_{j\to k}/N_j] = N_j[ p(k|j,T^*) \pm \delta ]$ i.e., there is some place for error. However, this error has a contribution that vanishes as $1/\tau$. For instance, we can look at this ``Bernoulli trial'' from a frequentist point of view. It is known that the MLE for $p(k|j,T^*)$ is indeed $N_{j\to k}/N_j$, and the error given by the inverse of the Fisher information is $\sqrt{(p(k|j,T^*) (1-p(k|j,T^*)))/N_j}$. By noting that $p(j|j,T^*)\propto \delta t$ and $N_j \propto \tau/\delta t$, we see that the error, compared to the main term vanishes as $1/\tau$.}. 
As a result we have
\begin{align}\label{eq:ln_P_T_nu}
    \ln \rho(T|\nu_N) = \ln \rho(T) + \sum_{j=0}^{1} {N_0 p(j|0,T^*)} \ln p(j|0,T) + \sum_{k=0}^{1} {N_1 p(k|1,T^*)} \ln p(k|1,T) + {\rm const.}
\end{align}

The rest of the proof is similar to the standard Bernstein-Von Mises theorem. Following \cite{Bernstein}, let the posterior be peaked around some value ${\tilde T}_{\rm MP}$. By Taylor expanding $\ln \rho(T|\nu)$ around this value and then taking its exponent, one gets
\begin{align}\label{eq:I_nu_def}
    \rho(T|\nu_N) & \propto \exp\left[ \ln \rho({\tilde T}_{\rm MP}|\nu_N) + \sum_{k\geqslant 2} \frac{d^k \ln \rho(T|\nu_N)}{k! d T ^k}|_{{\tilde T}_{\rm MP}}({{\tilde T}_{\rm MP}}-T)^k \right]\nonumber\\
    & \propto \exp\left[ -\frac{I(\nu_N)}{2}({{\tilde T}_{\rm MP}}-T)^2 \right] \prod_{k>2}\exp\left[ \frac{d^k \ln \rho(T|\nu_N)}{k! d T ^k}|_{{\tilde T}_{\rm MP}}({{\tilde T}_{\rm MP}}-T)^k \right],
\end{align}
where the first order expansion is zero because $\frac{d \ln \rho(T|\nu_N)}{ d T }|_{{\tilde T}_{\rm MP}}=0$---by definition of the maximum. In the second line we used that the zeroth order $\ln \rho({{\tilde T}_{\rm MP}}|\nu_N)$ is just a normalisation factor.
We also defined
\begin{align}
    I(\nu_N) \coloneqq -\frac{d^2 \ln \rho(T|\nu_N)}{ d T ^2}|_{{\tilde T}_{\rm MP}}.
\end{align}
In what follows we would like to (i) connect this term to the Fisher information, and (ii) to connect ${\tilde T}_{\rm MP}$ to the true temperature.
To this aim note that
by taking the first derivative of Eq.~\eqref{eq:ln_P_T_nu} and evaluating at the true temperature we have
\begin{align}
	\frac{d \ln \rho(T|\nu_N)}{dT}|_{T^*} 
	& = \frac{d \ln \rho(T)}{ d T }|_{T^*} + \sum_{j=0}^{1} {N_0 p(j|0,T^*)} \frac{d\ln p(j|0,T)}{dT}|_{T^*} + \sum_{k=0}^{1} {N_1 p(k|1,T^*)} \frac{d\ln p(k|1,T)}{dT}|_{T^*} \nonumber\\
	& = \frac{d \ln \rho(T)}{ d T }|_{T^*} + N_0\sum_{j=0}^{1} \frac{d p(j|0,T)}{dT}|_{T^*} + N_1\sum_{k=0}^{1}  \frac{d p(k|1,T)}{dT}|_{T^*} = \frac{d \ln \rho(T)}{ d T }|_{T^*},
\end{align}
which vanishes for a prior that has its maximum at the true temperature. This is always the case for the flat prior, which is our focus. As a consequence, the posterior's maximum is actually at the true temperature, i.e., ${{\tilde T}_{\rm MP}} = T^*$. 

Now, we also note that by using Eq.~\eqref{eq:ln_P_T_nu} one gets
\begin{align}\label{eq:I_T_tilde}
    I(\nu_N) = -\frac{d^2 \ln \rho(T)}{ d T^2 }|_{{\tilde T}_{\rm MP}} - \sum_{j=0}^{1} {N_0 p(j|0,T^*)} \frac{d^2\ln p(j|0,T)}{dT^2}|_{{\tilde T}_{\rm MP}} - \sum_{k=0}^{1} {N_1 p(k|1,T^*)} \frac{d^2\ln p(k|1,T)}{dT^2}|_{{\tilde T}_{\rm MP}}.
\end{align}
which by setting ${{\tilde T}_{\rm MP}} = T^*$ reduces to
\begin{align}
	I(\nu_N)  & = -\frac{d^2 \ln \rho(T)}{ d T^2 }|_{T^*} - \sum_{j=0}^{1} {N_0 p(j|0,T^*)} \frac{d^2\ln p(j|0,T)}{dT^2}|_{T^*} - \sum_{k=0}^{1} {N_1 p(k|1,T^*)} \frac{d^2\ln p(k|1,T)}{dT^2}|_{T^*}\nonumber\\
	& = -\frac{d^2 \ln \rho(T)}{ d T^2 }|_{T^*} - N_0 {F}[p(j|0,T^*)] -  N_1{F}[p(k|1,T^*)].
\end{align}
Now, as a final step, we have to replace for $N_0$ ($N_1$) according to their asymptotic values. This is given by the probabiity of thermal occupation of the ground (exited) state multiplied by the total number of measurements. We have
\begin{align}
	N_0 = Np_0(T) = N\frac{\Gamma_{\rm out}}{\Gamma_{\rm in}+\Gamma_{\rm out}},~~~N_1=N-N_1 = N\frac{\Gamma_{\rm in}}{\Gamma_{\rm in}+\Gamma_{\rm out}}.
\end{align}  
By using Eq.~\eqref{eq:fishercond} we get
\begin{align}
	I(\nu_N) & = -\frac{d^2 \ln \rho(T)}{ d T^2 }|_{T^*}
	 - \frac{N \delta t \Gamma_{\rm out}}{\Gamma_{\rm in}+\Gamma_{\rm out}} \left(\frac{\Gamma_{\rm in}'}{\Gamma_{\rm in}}\right)^2\Gamma_{\rm in} 
	 - \frac{N \delta t \Gamma_{\rm in}}{\Gamma_{\rm in}+\Gamma_{\rm out}}\left(\frac{\Gamma_{\rm out}'}{\Gamma_{\rm out}}\right)^2\Gamma_{\rm out}\nonumber\\
	& \overset{N\delta t = \tau}{=} -\frac{d^2 \ln \rho(T)}{ d T^2 }|_{T^*} 
	- \tau \frac{\Gamma_{\rm out}\Gamma_{\rm in}}{\Gamma_{\rm in}+\Gamma_{\rm out}}
	\left[ \left(\frac{\Gamma_{\rm in}'}{\Gamma_{\rm in}}\right)^2 + \left(\frac{\Gamma_{\rm out}'}{\Gamma_{\rm out}}\right)^2 \right]\eqqcolon I(\nu_{\tau}) ,
\end{align}
where we transit to the continious limit.
For long enough time $\tau$ this expression reduces to the Fisher information of the trajectory Eq.~\eqref{eq:Fisherrates}. Therefore, we prove that the long-time second order moment of the posterior distribution is given by the Fisher information. That is
\begin{align}
	\rho(T|\nu_{\tau}) \propto \exp\left[ -\frac{F[\rho(\nu_{\tau}|T^*)]}{2}({T^*}-T)^2 \right] \prod_{k>2}\exp\left[ \frac{d^k \ln \rho(T|\nu_{\tau})}{k! d T ^k}|_{T^*}(T^*-T)^k \right].
\end{align}
Finally, note that since $F[\rho(\nu_{\tau}|T^*)] \propto \tau$, we have the width of the posterior is $\propto 1/\sqrt{\tau}$. For hypothesis temperatures that fall within this range (i.e., $|T^*-T|~ < 1/\sqrt{\tau}$) we can ignore the term with $\prod_{k>2}$, since it involves at most terms that are of $\tau^{1-k/2}$ order. Then, by normalising the Gaussian distribution, we have
\begin{align}
	\rho(T|\nu_{\tau}) = \sqrt{\frac{F[\rho(\nu_{\tau}|T^*)]}{2\pi}} \exp\left[ -\frac{F[\rho(\nu_{\tau}|T^*)]}{2}({T^*}-T)^2 \right] .
\end{align}
\section{Other error quantifiers and estimators}\label{App:Distances&Estimators}
\begin{figure*}\label{fig:MAP_vs_MAP_app_vs_MLE}
    \includegraphics[width=.45\linewidth]{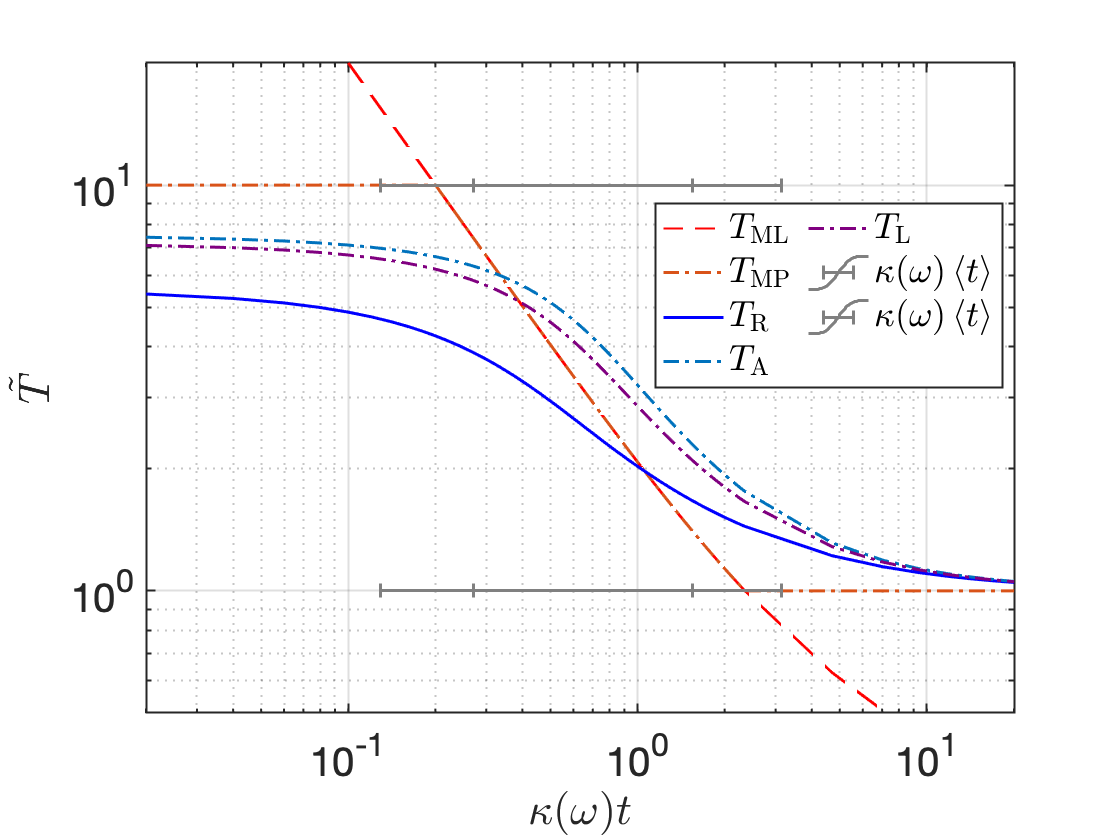}
    \includegraphics[width=.45\linewidth]{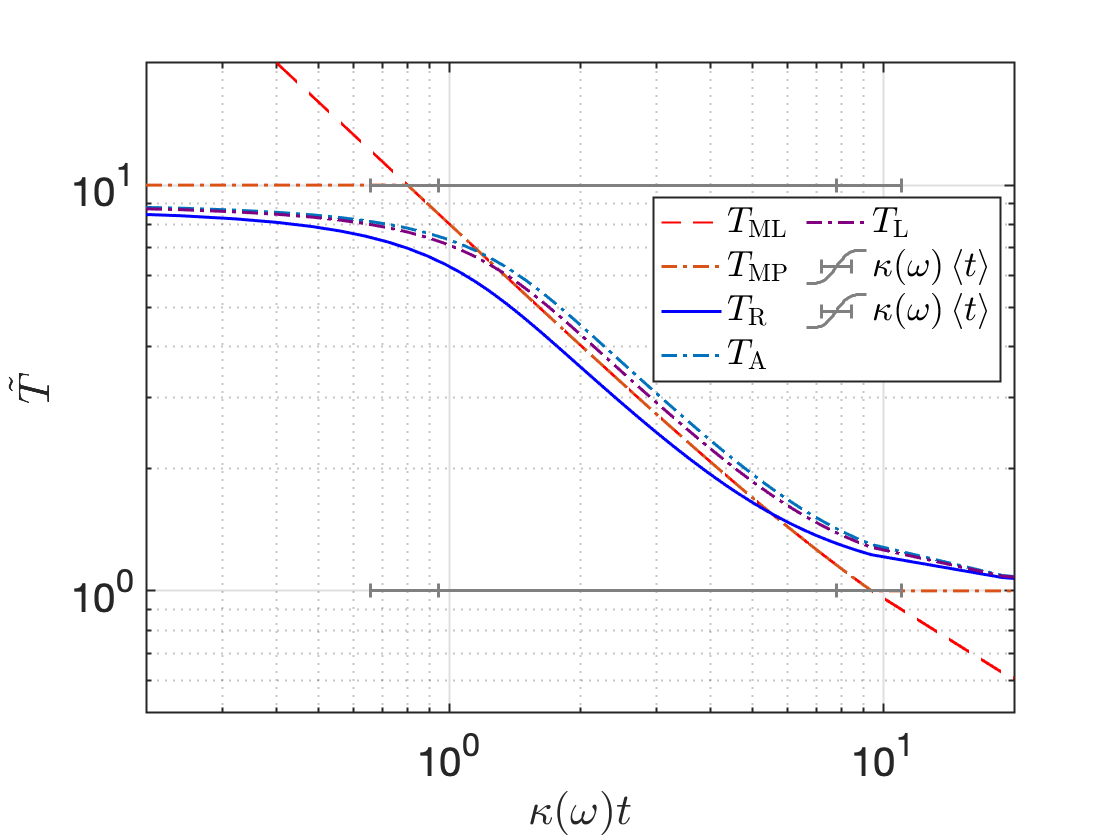}\\
    \includegraphics[width=.45\linewidth]{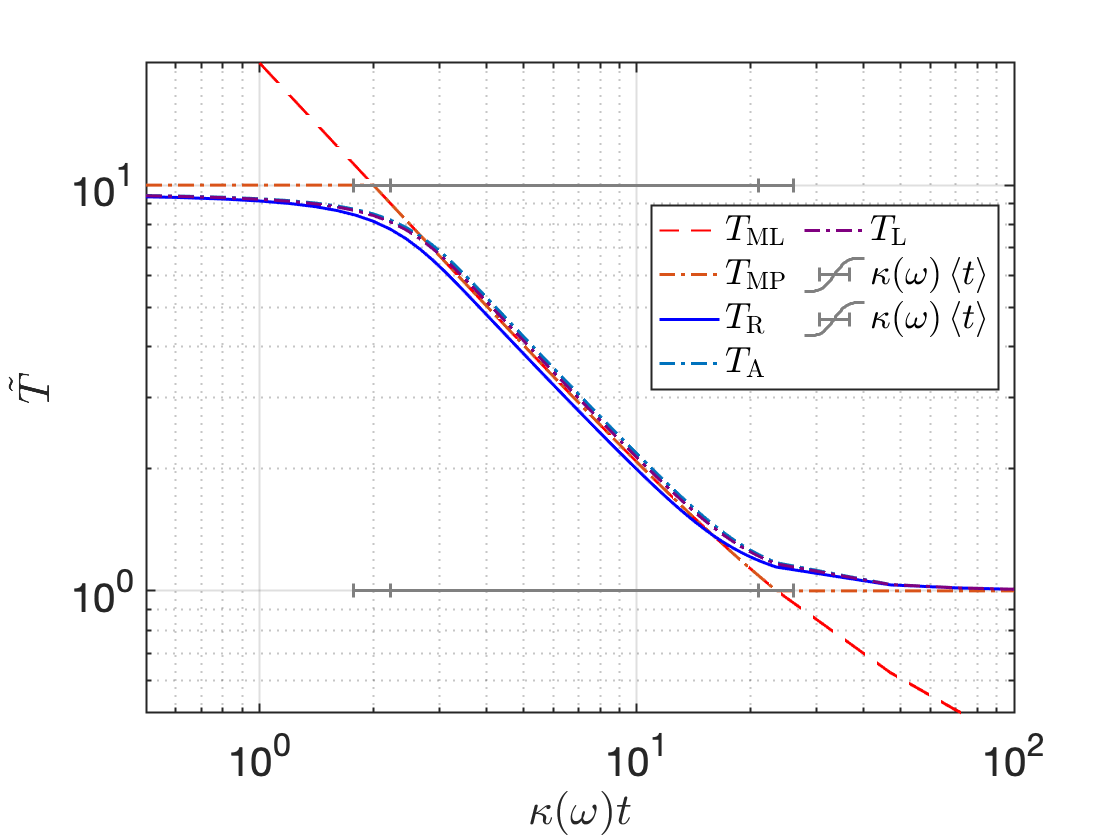}
    \includegraphics[width=.45\linewidth]{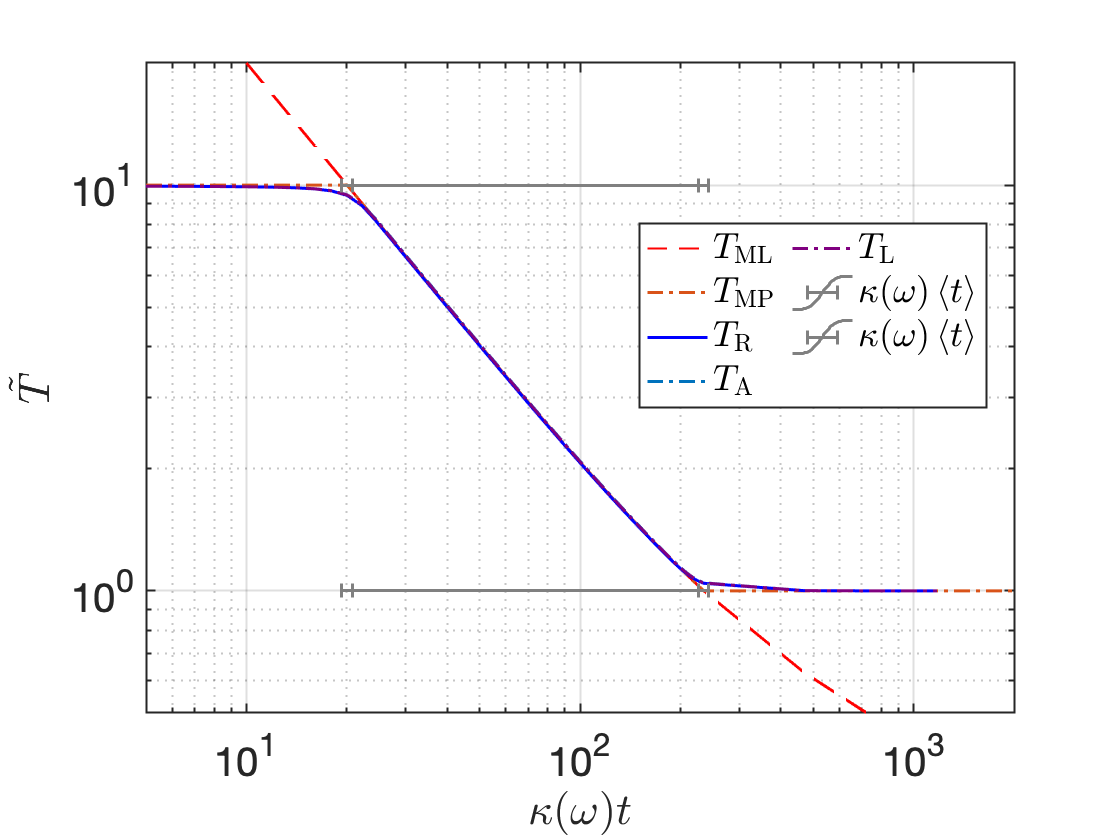}
    \caption{Illustration of the temperature that different estimators assign to the same observed trajectory. Suppose that we have a flat prior for temperature between $T_{\min}=1$ and $T_{\max}=10$. 
    The above graphs depict the temperature estimate via different estimators. 
    The gray graphs show the average time that it takes to make $m_{12}$ jumps up and $m_{21}$ jumps down, for different temperatures in the prior. The y-axis of the gray lines is not meaningful, it's repeated twice to help understanding the borders of the prior better. The error bars are found as the standard deviation of the total time. The gray lines shed light on when the ML and MP estimators are good enough. For e.g., the bottom right graph shows that at $\gamma_0 \kappa(\omega) t$ the relative estimator ${\tilde T}_{\rm R}$ deviates a lot from the ML. However, in the lab it is unlikely to take this long to observe $100$ jumps, because it falls outside the error bars of the total time. 
    The parameters are set to: top left $k=l=1$, top right $k=l=4$, bottom left $k=l=10$, and bottom right $k=l=100$.}
\end{figure*}

In the main text we took the relative distance as our cost function. However, our analysis is not restricted to this choice, and can be straightforwardly adopted for other cost functions. In particular, for the absolute mean square distance, the mean of the posterior is the optimal estimator: 
\begin{align}
        D_{\rm A}[{\tilde T}] &\coloneqq \int d\nu_{\tau} \rho(\nu_{\tau})\int dT \rho(T|\nu_\tau) \left({\tilde T}(\nu_\tau)-T\right)^2,\nonumber\\
        {\tilde T}_{\rm A}(\nu_\tau) &\coloneqq \int dT~T\rho(T|\nu_\tau),\nonumber
\end{align}
Moreover, for the mean square logarithmic error the optimal estimator reads~\cite{PhysRevLett.127.190402}
\begin{align}
        D_{\rm L}[{\tilde T}] &\coloneqq \int d\nu_{\tau} \rho(\nu_{\tau})\int dT \rho(T|\nu_\tau) \log^2\left[\frac{{\tilde T}(\nu_\tau)}{T}\right],\nonumber\\
        {\tilde T}_{\rm L}(\nu_{\tau}) &\coloneqq \exp\left[ \int dT\rho(T|\nu_\tau)\log T \right].\nonumber
\end{align}
Note that, in evaluation of the Bayesian estimator at long times, one often encounters exponentially small numbers due to the form of Eq.~\eqref{eq:probtraj}. Hence, to avoid numerical errors, one should use very precise arithmetic (e.g., in MATLAB), which slows down the numerics. However, as we see below, at long times and with large number of jumps, the Bayesian estimator can be very well approximated by the maximum likelihood estimator, which is much simpler to evaluate.

For a performance benchmark of the different estimators we discussed, notice that they are all only functions of the total number of jumps $k$, $l$, and the total time $\tau$. In particular, they are not affected by how long it takes before each individual jump happens. This allows us to compare how the different estimators perform, merely based on these parameters. Figure~\ref{fig:MAP_vs_MAP_app_vs_MLE} depicts an exemplary comparison. It shows that, for many jumps and long times, the ML and MP estimators (which can be given analytically) are very close to the optimal Bayesian estimators; hence one can use them as a good approximation in these situations, speeding up numerical simulations.

\section{Continuous measurement described using non-linear filtering}\label{app:KS_eqn}
The evolution of the prior under continuous measurements with Gaussian noise and a finite bandwidth can also be derived using Ito-stochastic calculus. In the non-noisy case, the population of the ground and excited state is monitored. This is equivalent to monitoring the difference in the probability to occupy the ground and excited states $\langle z \rangle = 2p_0(t)-1=\{ -1,1\}$. However, when the measurement is noisy the probability that the system is in the ground state can take any value between zero and one. Then if $r$ is the continuous signal from a noisy measurement with an infinite bandwidth we can define a Wiener increment
\begin{equation}
    dW= 2\sqrt{\lambda}(r- \langle z \rangle)dt.
\end{equation}
Introducing a finite bandwidth $\gamma$ the outcome $D(t)$ read out on the detector at time $t$ is related to the infinite bandwidth signal by, 
\begin{equation}
    D(t)=\int_{- \infty}^{t}ds \: \gamma e^{-\gamma(t-s) t}r(s)
\end{equation}

The observed outcome from the experiment is the solution to the stochastic differential equation
\begin{equation}\label{eq:meas_reg}
    dD=\gamma(2p_0(t)-1-D)dt+\\\frac{\gamma}{2\sqrt{\lambda}}dW.
\end{equation} 
By expanding the likelihood $P(dD|T)$ (Eq.~\ref{eq:meas_evol}) to first order in $dt$ and normalising the probability distribution we obtain the evolution of the ground state population~\cite{Annby-Andersson-PRL-2022},
\begin{align}
      dp_0(t)=&(\Gamma_{out}p_1(t)-\Gamma_{in} p_0(t)+ 8 \lambda p_0(t)p_1(t)(2p_0(t)-D-1))dt \\\nonumber
      &+\frac{8\lambda}{\gamma}p_1(t)dD. 
\end{align}

We now have enough information to derive the evolution of the state of knowledge about $T$. The change in this state of knowledge after an infinitesimal time step of the experiment is
\begin{equation}
 dP(T,t|\nu_{\tau}) =P(T,t+dt|\nu_{\tau},dD)-P(T,t|\nu_{\tau}).
\end{equation}
The evolution can then be broken up into an instantaneous part due to measurement and the time evolution after measurement
\begin{align}
dP(T,t|\nu_{\tau})=&P(T,t+dt|\nu_{\tau},dD)-P(T,t|\nu_{\tau},dD)\\ \nonumber
&+P(T,t|\nu_{\tau},dr)-P(T,t|\nu_{\tau}).
\end{align}
However, the state of knowledge will only evolve independently of measurements if predictions are made. Excluding prediction gives 
\begin{equation}\label{eq: diff}
 dP(T,t|\nu_{\tau}) =P(T,t|\nu_{\tau},dD)-P(T,t|\nu_{\tau}).
\end{equation} 
$P(T,t|\nu_{\tau},dD)$ can be computed using Bayes Rule:
\begin{align}
P(T,t|\nu_{\tau},dD) &=\frac{ P(dD|T)P(T,t|\nu_{\tau})}{\int dT P(dD|T)P(T,t|\nu_{\tau})}.
\end{align}

Then, to calculate the difference in Eq.\ref{eq: diff}, we first expand $P(T,t|\nu_{\tau},dD)/P(T,t|\nu_{\tau})$ to first order in $dt$ and to second order in $dD$ taking the expectation of the $dD^2$ terms since $\mathcal{E}(dD^2)=\gamma^2/(4\lambda)dt$.
That is
\begin{align}
    \frac{P(T,t|\nu_{\tau},dD)}{P(T,t|\nu_{\tau})}=&\frac{\text{exp}(\frac{-2\lambda}{\gamma^2dt}(dD-\gamma(2p_0(t)-D-1)dt)^2)}{\int dT P(dD|T)P(T,t|\nu_{\tau})}\\\nonumber
    =&\frac{\text{exp}(\frac{-2\lambda}{\gamma^2}(-2\gamma(2p_0-D-1)dD+\gamma^2(2p_0-D-1)^2dt))}{\int dT \text{exp}(\frac{-2\lambda}{\gamma^2}(-2\gamma(2p_0-D-1)dD+\gamma^2(2p_0-D-1)^2dt))P(T,t|D)}\\\nonumber
    =&1+\frac{4\lambda}{\gamma}((2\overline{p_0}^2-D-1-(2p_0-D-1)(2\overline{p_0}-D-1))dt\\\nonumber
    &+\frac{4\lambda}{\gamma}(2p_0-D-1-(2\overline{p_0}-D-1))dD +o(dt)
\end{align}
where, $ \overline{p_0}=\int dT p_0 P(T,t|\nu_{\tau}) $. Therefore, the evolution of the posterior is 
\begin{equation}\label{eq:evol_prior}
    dP(T,t|D)=\frac{8\lambda}{\gamma}(p_0-\overline{p_0})(dD-\gamma(2\overline{p_0}-D-1)dt). 
\end{equation}
The simulation of an experiment then consists of three parts. \begin{itemize}
    \item [(i)] First, the increment to the measurement register is simulated at each time step. This is done using the transition probabilities for the Poisson process and adding filtered white noise according to Eq.~\ref{eq:meas_reg}. This has to be done regardless of the choice of method for simulation.
    \item [(ii)] The measurement register is then used to solve the evolution of $p_0(t,T)$ at each temperature in the support. This is in contrast to the direct method outlined in the main body of the paper where Eq.~\ref{eq:meas_evol} is simulated.
    \item [(iii)] The prior is updated using the simulated measurement register and expectation values that were averaged over the previous increment of the prior. Once again this is different to the main text method where Eq.~\ref{eq:likelihood} is used to update  $P(T,t|\nu_{\tau})$.
\end{itemize} 

The simulation of the results in this way using the second order Milstein method leads to an improvement in calculation time by reducing the number of computational steps required in step (ii). This comes at the cost of introducing numerical instability when the time step is too large for a particular measurement strength. The Kushner-Stratonovich method is likely more sensitive to the size of the $dt$ increment because there are two expansions in $dt$ involved in simulating the evolution of the prior.
\section{More simulations of the noisy measurements}\label{app:bias}
\begin{figure}
    \centering
    \includegraphics[width=\linewidth]{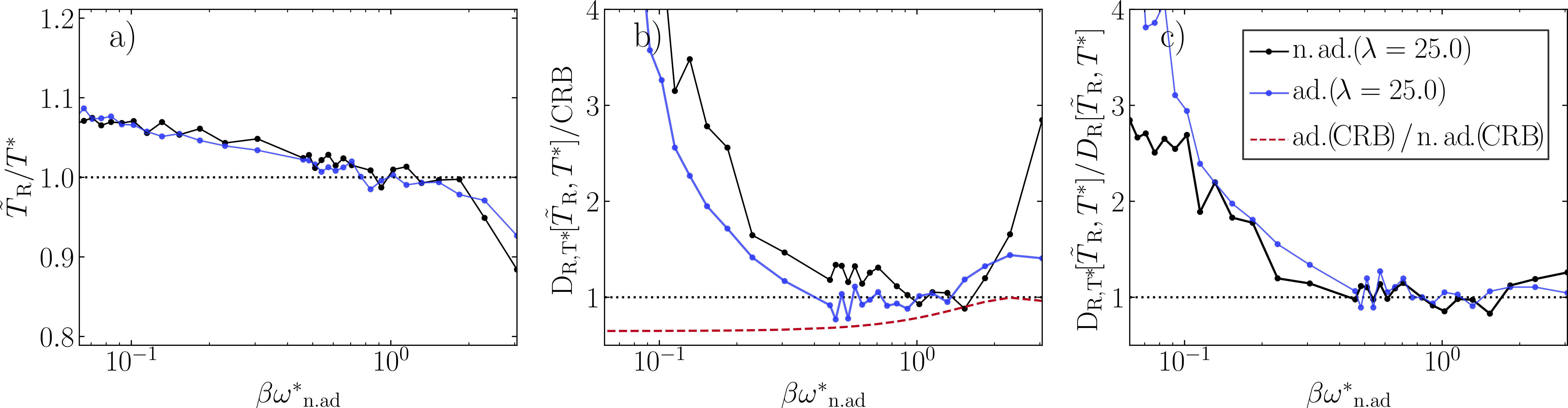}
    \caption{Comparison of the bias (a), relative error (b) and accuracy of the relative error (c) for noisy measurements with bandwidth $\gamma=10.0$ and different measurement strengths $\lambda$. The thermometry protocol was simulated for 35 true temperatures $T$ in the range $[0.1,10.0]\epsilon$ and with initial gap $\omega^*_{\text{n. ad}}$. At each of the temperatures the results are averaged over a set of 200 trajectories the points are joined with lines to guide the eye. The results are plotted at time $\kappa ' \epsilon\tau =100.0$. The shaded regions depict the standard deviation of the estimated temperature, true relative error and estimated error relative to the true error for each of the sets of trajectories. The black points and shaded regions correspond to the non-adptive strategy with with measurement strength $\lambda=25$. The blue points and shaded region represents the results for the adaptive strategy when the measurement strength is $\lambda=25$, here new trajectories need to be generated because the gap of the two level system changes in each step. The strength of the measurement has the greatest effect in reducing the bias and the bias is also the main contribution to the relative error. }
    \label{fig:bias}
\end{figure}
The noisy measurement scenario gives rise to a bias at low and high temperatures which prevents the scaling of the error from reaching the CRB. In this appendix the temperature dependence of the thermometry scheme is investigated in more detail. First, the ratio of the estimated and true temperature is shown in Fig.~\ref{fig:bias} (a). The estimated value is found for 35 different true temperatures in the range $[0.1,10.0]\epsilon$ and for each temperature the average estimate of 200 trajectories is calculated at the final time of each trajectory $\kappa' \epsilon \tau =100$. At high temperatures this ratio is larger than one meaning the temperature is over estimated and at low temperatures opposite behaviour is observed. An increase of the measurement strength causes the ratio to approach one for high temperatures but this is not observed at low temperatures. The adaptive strategy also improves the bias slightly for both high and low temperatures. 

Similarly, the true relative error of the adaptive and non-adaptive strategy is plotted relative to the non-adaptive CRB in Fig.~\ref{fig:bias} (b). Additionally, the ratio of the adaptive to non-adaptive CRB is plotted for reference. Agreement with this line would be the best an adaptive strategy could do. Here we see again that high temperatures are not able to reach the CRB but for a small window of temperatures the ratio approaches one. For the adaptive strategy, there is a large improvement when there is a large bias and for some temperatures the accuracy can go below the non-adptive CRB.

The right-most panel of Fig.~\ref{fig:bias} shows that the bias leads to the estimated error ${\rm D}_{\rm R}[\tilde T]$ under estimating the true error ${\rm D}_{{\rm R},T^*}[\tilde T]$. This ratio will approach one for more trajectories in the average according to Eq.~\ref{eq:freq_av}. This is important from an experimental point of view as it shows that in a temperature range where the bias is the main contribution to the error, the actual accuracy of the temperature estimate is lower than would be expected from the variance of the posterior when not enough iterations of the experiment are averaged over. Thus, at short times or for weak measurements, care must be taken to choose an appropriate temperature range for which the thermometer is reasonably accurate.
\end{document}